\documentclass[]{aastex631}
\usepackage{url}
\usepackage[acronym]{glossaries}
\usepackage{mathtools}

\makeglossaries

\newacronym{pB}{pB}{polarized brightness}

\newacronym{psi}{PSI}{Predictive Science Inc.}

\newacronym{mas}{MAS}{Magnetohydrodynamic Algorithm outside a Sphere Model}

\newacronym{mlso}{MLSO}{Mauna Loa Solar Observatory}

\newacronym{k-cor}{K-Cor}{K-Coronagraph Instrument at \acrlong{mlso}}

\newacronym{cor-1}{COR-1}{\acrshort{stereo} COR-1 Coronagraph Instrument}

\newacronym{pos}{POS}{plane-of-sky}

\newacronym{jsd}{JSD}{Jensen-Shannon Divergence}

\newacronym{kld}{KLD}{Kullback-Leibler Divergence}

\newacronym{mhd}{MHD}{Magnetohydrodynamic}

\newacronym{wcs}{WCS}{World Coordinate System}

\newacronym{pfss}{PFSS}{Potential-Field Source Surface}

\newacronym{qraft}{QRaFT}{Quasi-Radial Feature Tracing Algorithm \citep{vadim_uritsky_2022_7410948, uritsky2024}}

\newacronym{wsa}{WSA}{Wang-Sheely-Arge Model}

\newacronym{gsm}{GSM}{Geocentric Solar Magnetospheric}

\newacronym{fits}{FITS}{Flexible Image Transport System}

\newacronym{ne}{$n_e$}{electron density}

\newacronym{ne central}{\acrshort{mas} \acrshort{ne} \acrshort{pos}}{\acrshort{mas} central \acrlong{pos} \acrlong{ne}}

\newacronym{ne los}{\acrshort{mas} \acrshort{ne} \acrshort{los}}{ \acrshort{mas} \acrlong{los} integrated \acrlong{ne}}

\newacronym{los}{LOS}{line-of-sight}

\newacronym{forward}{FORWARD}{FORWARD: A Toolset for Model-Data Comparison \citep{2016FrASS...3....8G}}

\newacronym{forward pB}{\acrshort{forward} \acrshort{pB}}{\acrshort{forward} \acrlong{pB}}

\newacronym{cor-1 pB}{\acrshort{cor-1} \acrshort{pB}}{\acrshort{cor-1} \acrlong{pB}}

\newacronym{B}{$\vec{B}$}{magnetic field}

\newacronym{pos B}{\acrshort{pos} \acrshort{B}}{\acrlong{pos} \acrlong{B}}

\newacronym{hsd}{HSD}{Tukey's Honestly Significant Difference Test}

\newacronym{stereo}{STEREO}{Solar Terrestrial Relations Observatory}

\newacronym{stereo a}{\acrshort{stereo} A}{\acrshort{stereo} Ahead}

\newacronym{stereo b}{\acrshort{stereo} B}{\acrshort{stereo} Behind}

\newacronym{nasa}{NASA}{National Aeronautics and Space Administration}

\newacronym{toa}{ToA}{time of arrival}

\newacronym{cr}{CR}{Carrington rotation}

\newacronym{wtd}{WTD}{Wave-Driven-Turbulence}

\newacronym{imf}{IMF}{Interplanetary Magnetic Field}

\newacronym{em}{EM}{electromagnetic}

\newacronym{Q1}{Q1}{lower quartile}

\newacronym{Q3}{Q3}{upper quartile}

\newacronym{IQR}{IQR}{interqu\textcolor{black}{a}rtile range}

\newacronym{H0}{$H_0$}{null hypothesis}

\newacronym{ci}{CI}{confidence interval}

\newacronym{nsf}{NSF}{National Science Foundation}

\newacronym{roses}{ROSES}{Research Opportunities in Space and Earth Science}

\newacronym{cme}{CME}{coronal mass ejection}

\newacronym{angular difference}{\textcolor{black}{$\Delta\theta$}}{\textcolor{black}{angular difference}}

\newacronym{absolute angular difference}{$\left| \Delta\theta\right|$}{absolute \acrlong{angular difference}}

\newacronym{mean angular difference}{\textcolor{black}{$\overline{\left|\Delta\theta\right|}$}}{\textcolor{black}{mean \acrlong{absolute angular difference}}}

\newacronym{coronametric}{CoronaMETRIC}{Corona Magnetic-field Evaluation Through Real Image-model Comparison \citep{christopher_rura_2025_14921310}}

\newcommand{\spike}[2]% #1 = size of spike, #2 = centered text
{\bgroup
  \sbox0{#2}%
  \rlap{\usebox0}%
  \hspace{0.5\wd0}%
  \makebox[0pt][c]{\rule[\dimexpr \ht0+1pt]{0.5pt}{#1}}% top spike
  \makebox[0pt][c]{\rule[\dimexpr -\dp0-#1-1pt]{0.5pt}{#1}}% bottom spike
  \hspace{0.5\wd0}%
\egroup}

\begin{document}

\title{Quantitative Image-Based Validation Framework for Assessing Global Coronal Magnetic Field Models}

\correspondingauthor{Christopher Rura}
\email{rura@cua.edu}

\author[0000-0002-4992-180X]{Christopher E. Rura}
\affiliation{The Catholic University of America \\
 620 Michigan Avenue NE \\ 
 Washington, DC 20064, USA}
\affiliation{\acrshort{nasa} Goddard Space Flight Center \\
Code 670, Greenbelt, MD 20771, USA}

\author[0000-0002-5871-6605]{Vadim M. Uritsky}
\affiliation{The Catholic University of America \\
 620 Michigan Avenue NE \\ 
 Washington, DC 20064, USA}
\affiliation{\acrshort{nasa} Goddard Space Flight Center \\
Code 670, Greenbelt, MD 20771, USA}

\author[0000-0001-9498-460X]{Shaela I. Jones}
\affiliation{The Catholic University of America \\
 620 Michigan Avenue NE \\ 
 Washington, DC 20064, USA}
\affiliation{\acrshort{nasa} Goddard Space Flight Center \\
Code 670, Greenbelt, MD 20771, USA}

\author[0000-0003-1759-4354]{Cooper Downs}
\affiliation{\acrlong{psi} \\
9990 Mesa Rim Road, Suite 170, San Diego, CA 92121, USA}

\author[0000-0001-9326-3448]{Charles Nickolos Arge}
\affiliation{\acrshort{nasa} Goddard Space Flight Center \\
Code 670, Greenbelt, MD 20771, USA}

\author[0000-0001-5207-9628]{Nathalia Alzate}
\affiliation{ADNET Systems, Inc. \\
Greenbelt MD 20771, USA}
\affiliation{\acrshort{nasa} Goddard Space Flight Center \\
Code 670, Greenbelt, MD 20771, USA}

\submitjournal{ApJ} % Will typeset "Submitted to ApJ"
\received{March 14, 2025}
\revised{November 13, 2025}
\accepted{November 17, 2025}
% \published{date}

\begin{abstract}

% AAS Journals, the Astrophysical Journal (ApJ), the
% Astrophysical Journal Letters (ApJL), the Astronomical Journal (AJ), and
% the Planetary Science Journal (PSJ) all have a 250 word limit for the 
% abstract\footnote{Abstracts for Research Notes of the American Astronomical 
% Society (RNAAS) are limited to 150 words}.  If you exceed this length the
% Editorial office will ask you to shorten it. This abstract has 161 words.
Coronagraph observations provide key information about the orientation of the Sun’s magnetic field. Previous studies used \textcolor{black}{various algorithms to segment} quasi-radial features in coronagraph images \textcolor{black}{and} approximate the\textcolor{black}{ir} local \acrshort{pos} geometry and orientation \textcolor{black}{which} can be used as input for optimizing and constraining coronal magnetic field models. We present a new framework that allows for further quantitative evaluations of image-based coronal segmentation methods against magnetic field models, and vice-versa. We compare quasi-radial features identified from \acrshort{qraft}, a global coronal feature tracing algorithm, in white-light coronagraph images to outputs of \acrshort{mas}, an advanced \acrshort{mhd} model. We use the \acrshort{forward} toolset to produce synthetic \acrshort{pB} images co-aligned to real coronagraph observations, segment features in these images, and quantify the difference between the inferred and model magnetic field. This approach allows us to geometrically compare features segmented in artificial images to those segmented in white-light coronagraph observations against the \acrshort{pos} projected \acrshort{mas} coronal magnetic field. We quantify \acrshort{qraft}’s performance in the artificial images and observational data, and perform statistical analy\textcolor{black}{s}es that measure the accuracy and uncertainty of the model output to the observational data. The results demonstrate that a coronal segmentation method identifies the global large-scale orientation of the coronal magnetic field within $\sim\pm10^\circ$ of the \acrshort{pos} projected \acrshort{mas} magnetic field.

\end{abstract}


\section{Introduction}

\subsection{Overview}
\label{subsec:Introduction_overview}

The solar corona is an important region of the solar atmosphere that sets the conditions of space weather propagation into the Heliosphere. \textcolor{black}{In this region, the} coronal magnetic field is dominant, \textcolor{black}{and it is a} key \textcolor{black}{element of} most phenomena \citep{Solanki_2006}. \textcolor{black}{T}he magnetic field in the atmospheric layers above the photosphere is notoriously difficult to measure, and therefore we must rely on extrapolations from indirect measurements of the \textcolor{black}{photospheric} magnetic field to offer the most accurate understanding of the coronal magnetic field \citep{Solanki_2006}. For this reason, \textcolor{black}{assessing the validity of coronal models} is essential for accurate space weather forecasting.

% In this paper we present a method to evaluate an image based method for assessing coronal magnetic field models. This method was developed by \citet{jones2016,jones2020}, who presented a method to optimize magnetic field models by comparing the orientation of quasi-radial features segmented in \acrfull{pB} observations to the orientation of the magnetic field model.  This approach depends on the assumption that structure seen in coronal images can be treated as a proxy for the orientation of features in the coronal magnetic field. In this paper we will test the accuracy of this assumption.

Magnetic field line tracing methods use optical features observed in coronal images to segment large-scale signatures of the coronal magnetic field to measure its orientation. A major open problem in space physics is that there is little to no ground-truth precedent in which to compare the orientation of these segmented features to, which makes it difficult to determine if they are accurately representing magnetic structure. While magnetic field models such as \acrfull{pfss} and \acrfull{mhd} models can approximate the coronal magnetic geometry, it is challenging to constrain them due to the absence of magnetic field measurements in the corona. \textcolor{black}{One method to assess the coronal magnetic field is assuming that the geometry of structures seen in coronal images can be treated as a proxy for the magnetic field configuration. Based on this assumption,} \citet{jones2016,jones2020} \textcolor{black}{developed a method} comparing the orientation of quasi-radial features segmented in \acrfull{pB} observations to the orientation of the magnetic field model. \textcolor{black}{These} studies \textcolor{black}{were based on} the automated \acrfull{qraft}, which uses adaptive thresholds to enhance and segment coronal features and approximates their projected geometries. This algorithm focuses on tracing open field line geometries, which have very weak optical signatures and are typically difficult to measure.  The features observed in the images \textcolor{black}{are then assumed to} reflect the projected magnetic field in the image plane. 

% Evaluating and validating these methods against a ground-truth may illuminate to what extent we can rely on the features they produce.

In this study, we will test this assumption using numerical outputs of the \acrfull{mas} from \acrshort{psi}, which is described in \textcolor{black}{S}ection \ref{sec:mas_model_description}. This \acrshort{mhd} model is useful because it incorporates a complete energy equation, which is necessary in describing both the plasma and magnetic properties of the corona. We use the \textit{SSWIDL} package \acrshort{forward} to create synthetic white-light images based on the \acrshort{mas} model solution for the 2017 solar eclipse \citep{2018NatAs...2..913M}. These images will be as seen from the perspective of the \acrshort{stereo} satellite at times corresponding to the \textcolor{black}{white-light corona observations from} the \acrfull{cor-1}. These methods will be discussed in more detail in \textcolor{black}{S}ection \ref{sec:data}. We present our results in this analysis in \textcolor{black}{S}ection \ref{sec:results}, discuss potential errors in \textcolor{black}{S}ection \ref{sec:discussion}, and state our conclusions in \textcolor{black}{S}ection \ref{sec:conclusions}.

\begin{figure}
\noindent\includegraphics[width=\textwidth]{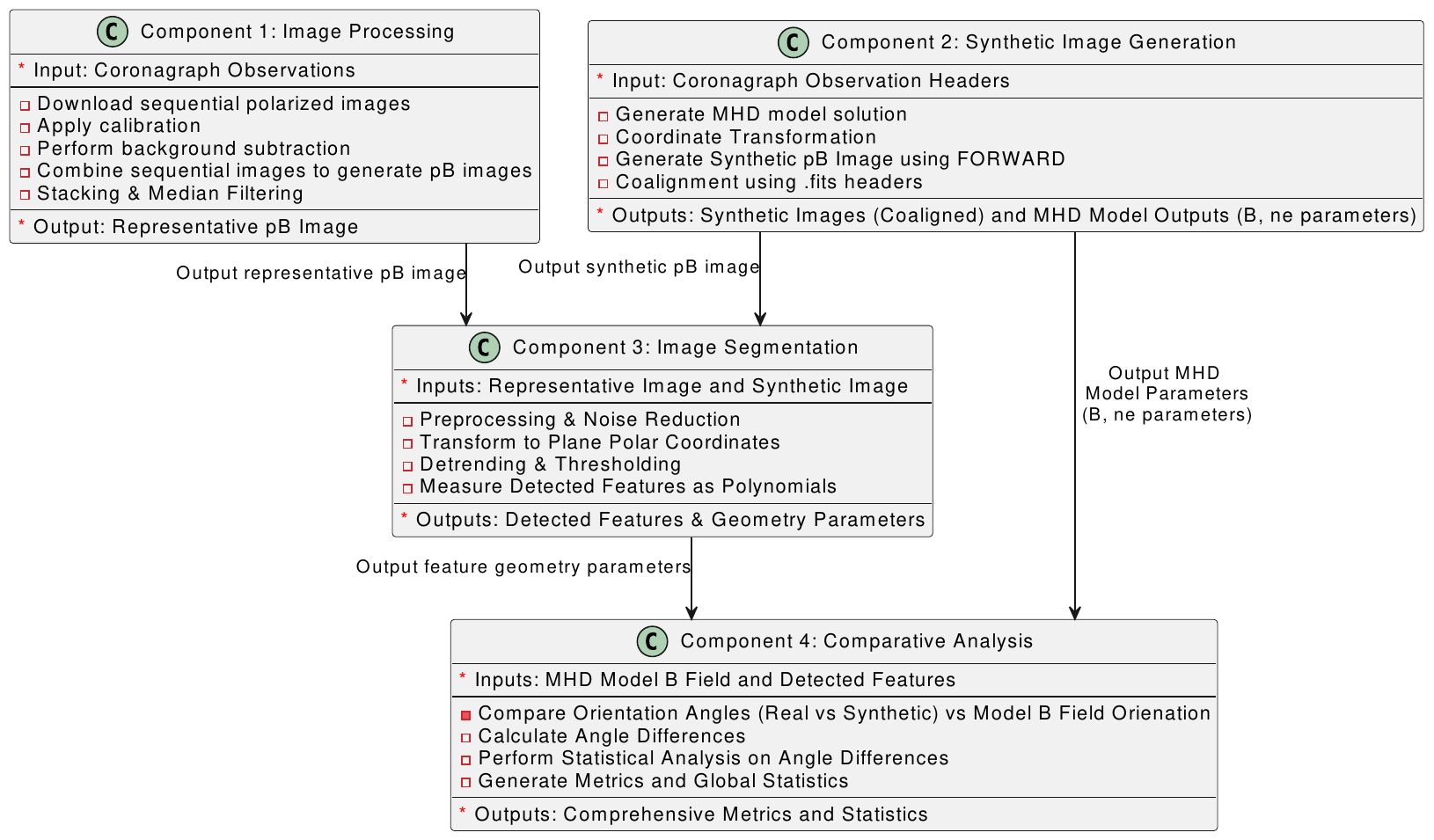}
\caption{Flowchart style overview of defined framework for quantitatively assessing image-based coronal segmentations against magnetic field models, and vice-versa. Each arrow represents an output from a component of the framework that is used as an input into the component the arrow is connected to.}
\label{fig:framework_overview}
\end{figure}

% These methods will be discussed in more detail in \textcolor{black}{S}ection \ref{sec:data}. By comparing the optical structure of quasi-radial features detected in coronal images to synthetic images generated by this \acrshort{mhd} model as well as its expected magnetic field, we seek to provide a unique evaluation of this method's ability to segment true magnetic structure. We present our results in this analysis in \textcolor{black}{S}ection \ref{sec:results}, discuss potential errors in \textcolor{black}{S}ection \ref{sec:discussion}, and state our conclusions in \textcolor{black}{S}ection \ref{sec:conclusions}. 

\textcolor{black}{This work also provides} a framework to evaluate image-based coronal segmentation methods against model magnetic fields and vice-versa quantitatively. A high level overview of this framework is shown in \textcolor{black}{F}igure \ref{fig:framework_overview}, emphasizing the generality of this approach, which allows for the evaluation of segmentations from multiple data sources and segmentation algorithms against multiple magnetic field models. The potential use of this framework will be further discussed in \textcolor{black}{S}ection \ref{sec:discussion}.

\subsection{Theoretical Basis}\label{subsec: k-corona}

The work of \citet{jones2020} used features segmented in \acrshort{pB} images, which highlight the K-corona. These features are a result of intensity gradients present in the \acrshort{pB} image, however this structure can be connected to the plasma density distribution in the K-corona.  The K-corona itself is the result of photons emitted from the photosphere scattering off of coronal free electrons \citep{1966gtsc.book.....B}.  This process is known as Thomson scattering. Since the electric field vector is normal to the direction of propagation for electromagnetic radiation, the K-Corona is polarized \citep{BLACKWELL19671}. As discussed in \citet{2009SSRv..147...31H}, the \acrshort{pB} of the white-light corona can be modeled by the integration of the electron density along an observer's line of sight. This integration must take into account several factors, such as the scattering function, the decreasing illumination of electrons with distance from the sun, and the decreasing degree of polarization with distance from the observer's \acrfull{pos}.

The frozen-in-flux condition states that the bulk motion of an ideal \acrshort{mhd} plasma cannot move perpendicularly to the \acrfull{B} without \textcolor{black}{dragging the field along}, or vice versa. The magnetic field therefore behaves as if it is 'frozen-in' with the plasma, so that the magnetic field moves with the plasma. This condition applies in highly conductive plasmas, such as in the corona and solar wind, as it accelerates into the Heliosphere \citep{2013MurrayPhDT}. Due to this, the plasma is intimately connected to the magnetic field in this region of the solar atmosphere, as the plasma evolution along individual flux tubes is isolated from neighboring flux tubes. The practical effect of this is that any density and temperature contrasts between adjacent flux tubes will naturally highlight the structure of the magnetic field itself \citep{2005psci.book.....A}. \textcolor{black}{Additionally, high density streamers and pseudostreamers are associated with the neutral current sheet. These structures are indicated by a high squashing factor $Q$, which may be a way to differentiate streamers from other radial features \citep{2021ApJ...909...10A}. Ultimately}, the structure seen in coronal \acrshort{pB} images should in theory serve as a proxy for the orientation of features in the coronal magnetic field.

\section{Data and Methodology} \label{sec:data}

\subsection{White-Light Observations}
\label{subsec:observations}

The white-light K-Corona is generated due to the electron scattering of photospheric light, meaning it is linearly polarized in a tangential direction to the solar limb. Coronagraphs take advantage of this by measuring the \acrfull{pB} of the corona, which is directly related to its electron density \citep{2012SPIE.8444E..3ND}. The \acrfull{cor-1} \textcolor{black}{onboard} \acrshort{nasa}'s \acrshort{stereo a} spacecraft measure\textcolor{black}{s} the \acrshort{pB} of the corona. Its design is described more completely in \citet{Thompson2003}, but in short, it is an occulting refractive coronagraph that utilizes an internal Lyot stop that is adapted to be used in space \citep{Thompson2010}.

\textcolor{black}{A}s mentioned in the first component of \textcolor{black}{F}igure \ref{fig:framework_overview}, images from this data source are downloaded at level 0 and processed to level 1 using the SolarSoft \textit{secchi\_prep} routine. This process involves background subtraction, vignetting, flat-field correction, CCD bias correction, and other calibration processing. This routine was run with the keywords \textit{polariz\_on}, \textit{pB}, and \textit{rotate\_on} to calculate the polarized brightness and rotate the image to solar north. This routine combines three polarized images, (at $0^{\circ}$, $120^{\circ}$, and $240^{\circ}$), taken in sequence into one \acrshort{pB} image. These images were produced at a 10 minute cadence over an 8-hour timespan to retrieve 50 \acrshort{pB} images. These images are then stacked and a median is \textcolor{black}{calculated} to create a representative image \textcolor{black}{with} enhance\textcolor{black}{d} signal and \textcolor{black}{reduced} noise. A similar process is described in \citet{Thompson2010}. This process is useful for identifying quasi-radial plasma density features \textcolor{black}{which}  remain static over a short period of time, allowing us to enhance the signal of these features and filter out imperfections such as cosmic rays and other image artifacts. Despite this enhancement, there are still artifacts present in the final output of this process, which will be discussed in further detail in \textcolor{black}{S}ection \ref{sec:discussion}.

\subsection{\acrshort{psi} \acrshort{mas} Model}

%[section to be added by Cooper]
\label{sec:mas_model_description}

To create a ground-truth testing framework we leverage outputs from state-of-the-art numerical models of the solar corona. Component two of \textcolor{black}{F}igure \ref{fig:framework_overview} showcases how these outputs are used in the high level overview of this framework. To this end we employ the \acrfull{mas} model of the global solar corona \citep{mikic99,lionello09,2018NatAs...2..913M}. This model \textcolor{black}{and similar codes} \citep[\textcolor{black}{e.g.,}][]{vanderholst14,reville20} employs a `thermodynamic' \acrshort{mhd} approach, where the additional terms that describe energy flow in the corona and solar wind are included, including coronal heating, parallel thermal conduction, radiative loss, and Alfv\'en wave acceleration. This treatment is essential for capturing the thermal-magnetic state of the corona, including the all-important interplay between magnetic and hydrodynamic forces that open the corona and form the solar wind.  Realistic magnetic configurations for targeted time-periods are obtained by using measurements of the photospheric magnetic field as the primary boundary condition. Most importantly, using such a model enables the direct computation and comparison of forward modeled observables to real observations. This technique can be used to both directly constrain the coronal model and connect coronal observables to their underlying physical state of the plasma \citep{lionello09,boe21,boe22}.

For this particular study we use the high-resolution MAS coronal prediction simulation developed for the August 21, 2017 total solar eclipse. This simulation is fully described by \citet{2018NatAs...2..913M}, but can be summarized as follows: The simulation utilizes a \acrfull{wtd} approach to specify coronal heating \citep{lionello14,downs16} and an energization technique to add field aligned currents (shear/twist) over large-scale polarity inversion lines \citep{2018NatAs...2..913M,yeates18_ssrv}. To model the corona at a given time, the model must specify a full-sun map of the radial component of the magnetic field at the inner boundary. Because of the lead-time required for publishing the prediction, this simulation used a \textcolor{black}{manually constructed} splice of synoptic map data from SDO/HMI \citep{scherrer12} based on what was available about 10 days prior to the eclipse. This map combined data for \acrfull{cr} 2192 with near-real-time data from a part of \acrshort{cr} 2193. To capture plume-like density structures at the poles, the polar caps were filled with a random flux distribution \textcolor{black}{matching} net flux \textcolor{black}{observations} at high latitudes. \textcolor{black}{T}he boundary magnetic field measurements span a time-range of approximately July 16 to August 11, 2017, which placed the oldest data near the west limb during totality. 

\subsection{Synthesizing \acrshort{mhd} Model Solution into Synthetic Images }

\label{subsec:synthesizing_synthetic_images}

 \textcolor{black}{W}e use the toolset \acrshort{forward} in our framework to transform the 3D \acrshort{mas} solution into \acrshort{pos} view using Carrington Heliographic coordinates, where the \acrshort{pos} is given by the Y and Z axes with the X-axis representing the Sun-observer line \citep{2016FrASS...3....8G}.
\textcolor{black}{T}he K Corona's \acrshort{pB} may be computed as the integration of the electron density along an observer's \acrfull{los}. \textcolor{black}{This integration must take into account several factors, such as the scattering function, the decreasing illumination of electrons with distance from the sun, and the decreasing degree of polarization with distance from the observer's \acrshort{pos}.} \citet{2009JGRA..114.6101S} provides an explicit equation to calculating the \acrshort{pB} in a \acrshort{mas} context. \acrshort{forward} utilizes a scattering function that depends on both the radial distance to the \textcolor{black}{p}hotosphere and the geometry of the observer relative to the scattering point to incorporate illumination of the K Corona due to Thomson scattering \citep{2016FrASS...3....8G}. This process results in a synthesized \acrshort{pB} image in the \acrshort{pos} view from the perspective of a hypothetical observer. These synthetic \acrshort{pB} images are then co-aligned to each respective observation by using \acrshort{wcs} coordinates extracted from each observation's \acrshort{fits} header as described in \citet{wcs_coalignment_tutorial}. Additionally, we estimate the radius of the occulting disk for each synthetic \acrshort{pB} image using parameters from each observation's \acrshort{fits} header. Figure \ref{fig:cor-1_pB_model_comparison} shows (\textcolor{black}{right}) a \acrlong{cor-1 pB} observation used in this analysis, taken on 2017-08-29, and (\textcolor{black}{left}) the synthetic \acrlong{pB} projected onto the \acrlong{pos} from the equivalent observational perspective with intensity displayed in a log scale.

\subsection{\acrshort{qraft} Methodology}
\label{subsec:qraft}

\textcolor{black}{W}e use the \acrshort{qraft} code as our image-based segmentation method of choice in our framework, (component three of \textcolor{black}{F}igure \ref{fig:framework_overview}). \acrshort{qraft} enables an automatic detection of quasi-radial coronal features, which are expected to be approximately aligned with the local coronal magnetic field. \textcolor{black}{\citet{uritsky2025} describes \acrshort{qraft}'s methodology more comprehensively, and serves as a companion paper to this analysis that focuses specifically on \acrshort{qraft}'s uncertainties and applications.} \textcolor{black}{H}ere we present a concise outline of the main processing steps.

The \textcolor{black}{input} coronal image $I(x,y)$ first undergoes a preprocessing procedure aimed to reduce the level of the high-frequency pixel noise and to mitigate the systematic decay of the image intensity with the distance from the solar limb. \textcolor{black}{T}he radial detrending is based on subtracting an azimuthally-averaged large-scale radial trend from the processed image. At the next step, the coronagraph image is transformed into plane polar coordinates $(\phi,\rho)$, with the position angle $\phi \in [0, 2\pi]$ measured in the counterclockwise direction relative to the western solar limb and the radial coordinate $\rho$ representing the \acrshort{pos} distance from the solar disk center. This coordinate system is a natural choice for studying quasi-radial coronal structures in open-flux solar regions since such structures are characterized by a slowly varying position angle. Other types of structures such as closed loops require a different approach and lie outside of the \acrshort{qraft}’s applicability domain. \textcolor{black}{The transformed polar-coordinate images had the angular resolution of $1.0^{\circ}$ and the radial resolution $21.8$ arcsec, with the image dimensions of $365 \times65$ pixels in the azimuthal and radial directions, respectively, (see Table 2 of \citet{uritsky2025} for the full set of processing parameters).}

To enhance the quasi-radial structures, we compute the unsigned second-order position-angle difference over a specified characteristic azimuthal scale $\Delta \phi$:

\begin{equation}
\Delta^2 I(\phi, \rho) = | I(\phi-\Delta \phi/2, \rho) + I(\phi+\Delta \phi/2, \rho) - 2 I(\phi, \rho) |
\label{eq_diff}
\end{equation}
We found that the second-order azimuthal difference to be a sensitive marker of field-aligned image structures, with the absolute value operator used in Eq. \ref{eq_diff} enabling a consistent tracing of the structure shape on either side of the structure. \textcolor{black}{The differencing scale $\Delta \phi$ has been chosen to be $2^\circ$ based on preliminary performance tests. See \citet{uritsky2025} for more details.}

The enhanced image (\ref{eq_diff}) is detrended in the position angle direction to improve the detection \textcolor{black}{of} structures \textcolor{black}{with} different brightness and signal-to-noise levels. The detrending is achieved by dividing the enhanced image by its low-pass filtered counterpart obtained by boxcar averaging the array $\Delta^2 I(\phi, \rho)$ along the $\phi$ direction. Periodic boundary conditions were applied at the edges of the $\phi$ interval for when performing the azimuthal differencing and detrending. 

The resulting enhanced image is subject to an adaptive thresholding allowing us to identify contiguous clusters of image pixels associated with quasi-radial coronal features of interest. The method 
``label region'' of the IDL language is used to identify and separate all clusters of pixels satisfying the chosen detection threshold. The cluster detection is performed multiple times by using different detection thresholds representing different percentile levels of the studied image, as well as by varying the lower bound $\rho_{min}$ of the processed range of radial positions to focus the algorithm on different coronal altitudes. We use between 10 and 20 percentile threshold choices and 10 $\rho_{min}$ values for processing each coronal image, which results in 100 - 200 individual tracing runs. Combining the outputs of these runs provides a relatively uniform detection performance at different locations around the Sun. 

\begin{figure*}
\begin{center}

PSI MAS / FORWARD pB Model \hspace{4.0 cm} COR-1 Observation 2017-08-29

\includegraphics[width=8.7 cm]{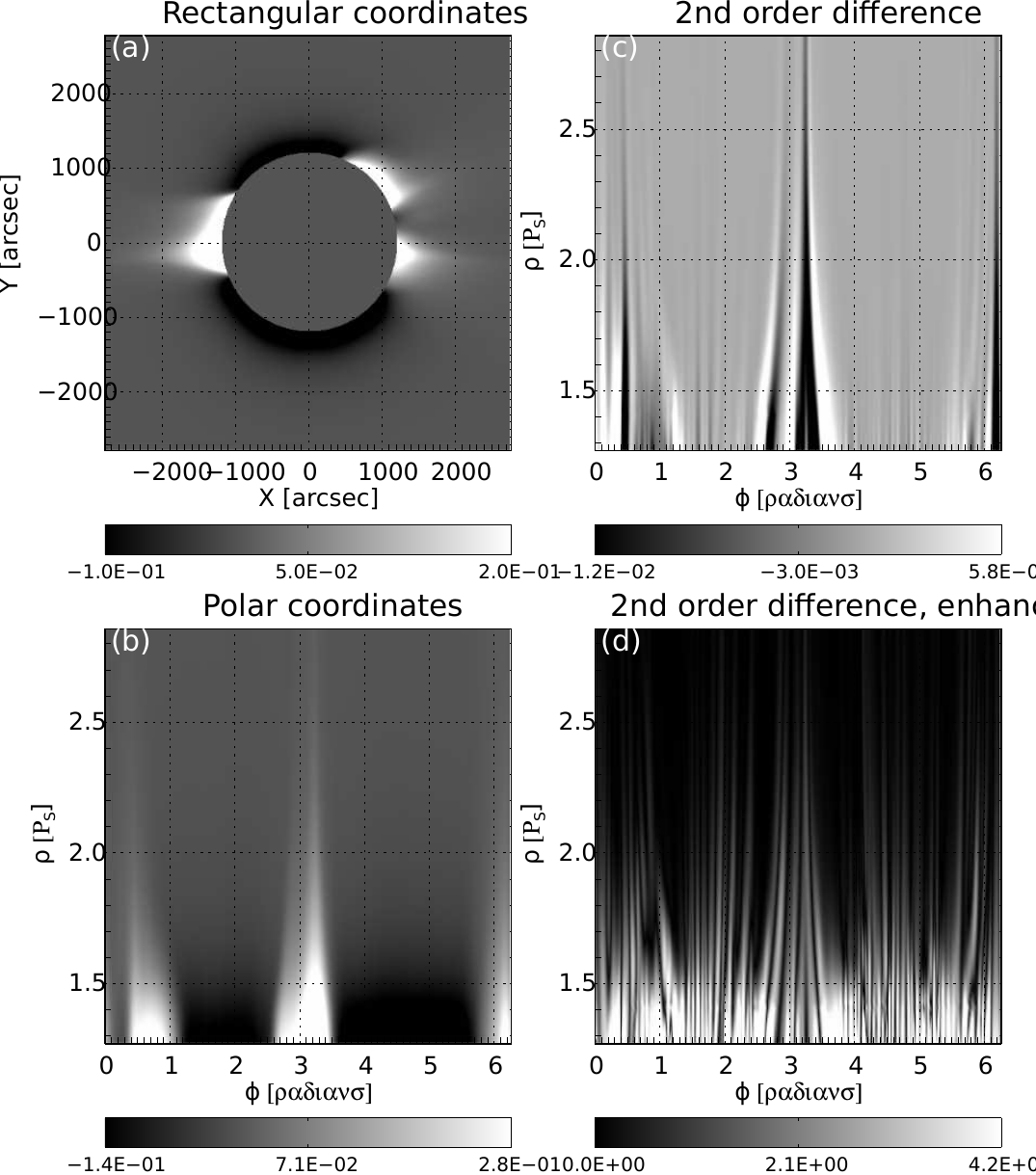} \,\,\,\, \includegraphics[width=8.7 cm]{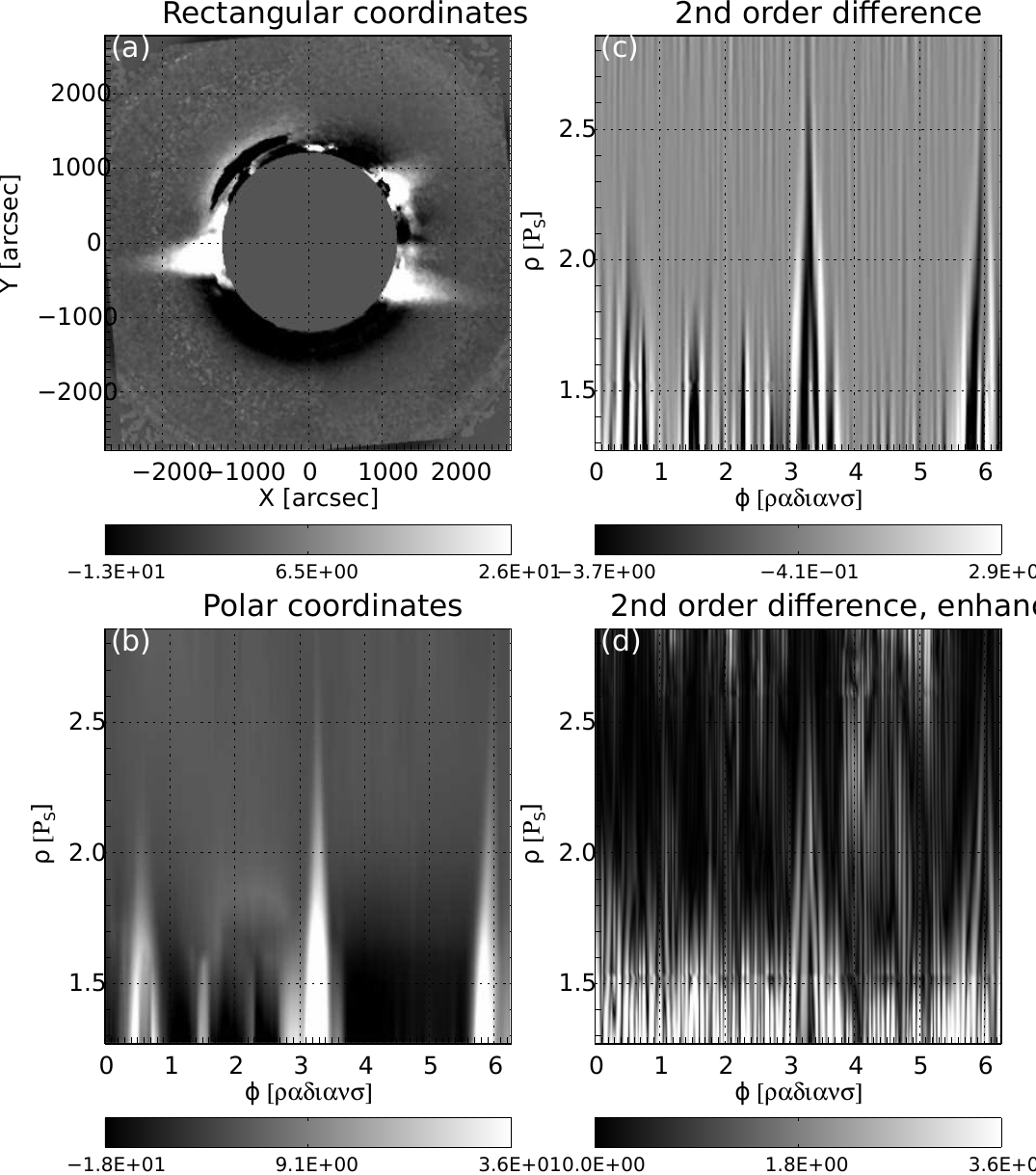}

\vspace{0.3cm}
\includegraphics[width=8.7 cm]{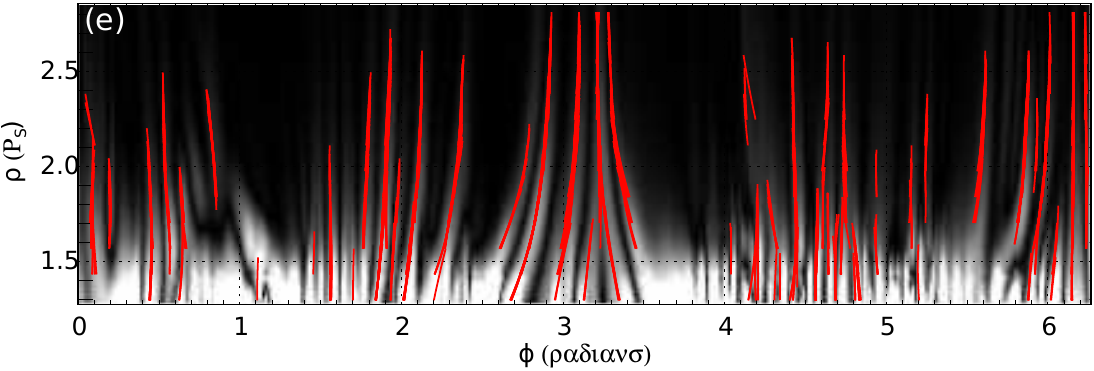} \,\,\,\, \includegraphics[width=8.7 cm]{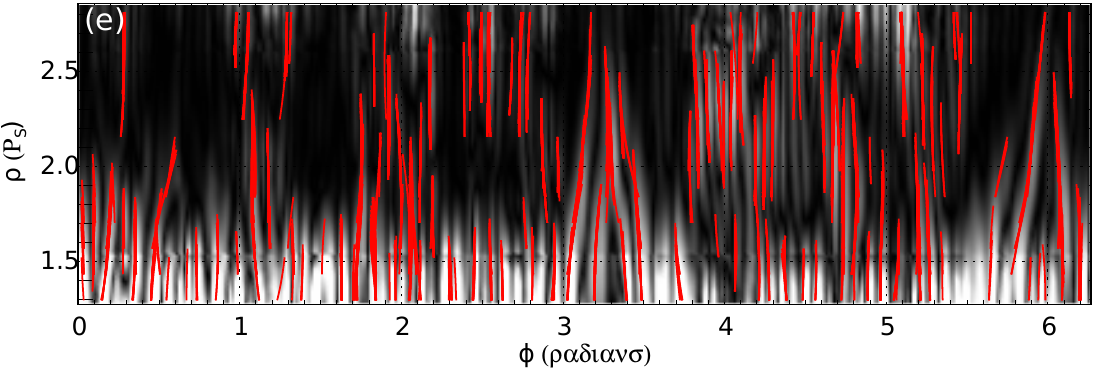}

\caption{\label{fig_qraft_example} 
\acrshort{qraft} processing steps shown for (left) a synthetic \acrfull{pB} image produced by \acrshort{mas} and the \acrshort{forward} code and (right) an observed \acrshort{pB} image obtained from \acrshort{cor-1} on 2017-08-29. The \acrshort{qraft} image is aligned with the \acrshort{cor-1} image as explained in the text. In each example the upper left panel (a) is the original \acrshort{cor-1 pB} image after radial detrending and smoothing, the bottom left panel (b) is the same image in plane polar coordinates, the top right panel (c) is the signed second-order azimuthal difference, the bottom right panel (d) is the unsigned second order difference after azimuthal detrending revealing a fine quasi-radial structure in the example observed image, and (e) are the resulting segmented features created by tracing the azimuthal gradients and automatically removing unrealistic features.}
\end{center}
\end{figure*}

The clusters produced by each tracing run are  approximated by polynomial functions using a nonlinear least-square fitting. These fits are used to calculate  the \acrshort{pos} orientation angles along the central line of each of the detected structures. Figure \ref{fig_qraft_example} shows an example of each of these processing steps on a \acrshort{cor-1 pB} observation used in this study\textcolor{black}{, as well as a synthetic \acrshort{pB} image produced by \acrshort{mas}.} \textcolor{black}{The detected image features have been approximated by second-order polynomials providing an analytical description of the $\phi (\rho)$ dependence for each feature. The polynomial models enabled a more robust analysis of the local geometry of the features, compared to a direct pixel by pixel analysis. A similar fitting methodology has been recently described by \citet{2023ApJ...945..116A, 2024ApJ...973..130A} who used a third-order polynomial representation of the edges of coronal streamers for tracking non-radial plasma outflows. Compared to their approach, our fitting algorithm focuses on quasi-stationary features observed over a broader range of coronal conditions, including those characterized by signal to noise ratios which are significantly lower than those found at streamer boundaries.} 

\textcolor{black}{The detected \acrshort{qraft} features have been validated using a set of automatic filters which allowed us to eliminate unreliable or irrelevant results. The filtering conditions included a verification of feature width, length, aspect ratio and curvature, to ensure that only sufficiently elongated field-aligned groups of pixels are included in the analysis, as well as the optical intensity of the features. See \citet{uritsky2025} for more details about the filtering process implemented in QRaFT.}

\subsection{Data/Model Comparison}
\label{subsec:data-model-comparison}

With the \acrshort{mas} model providing a ground truth magnetic field, we segment features with the \acrshort{qraft} algorithm, and measure how well their orientations match the model field orientation. By comparing the optical structure of quasi-radial features detected in coronal images to synthetic images generated by this \acrshort{mhd} model as well as its expected magnetic field, we seek to provide a unique evaluation of this method's ability to segment true magnetic structure.

\begin{figure}
\noindent\includegraphics[scale=0.5]{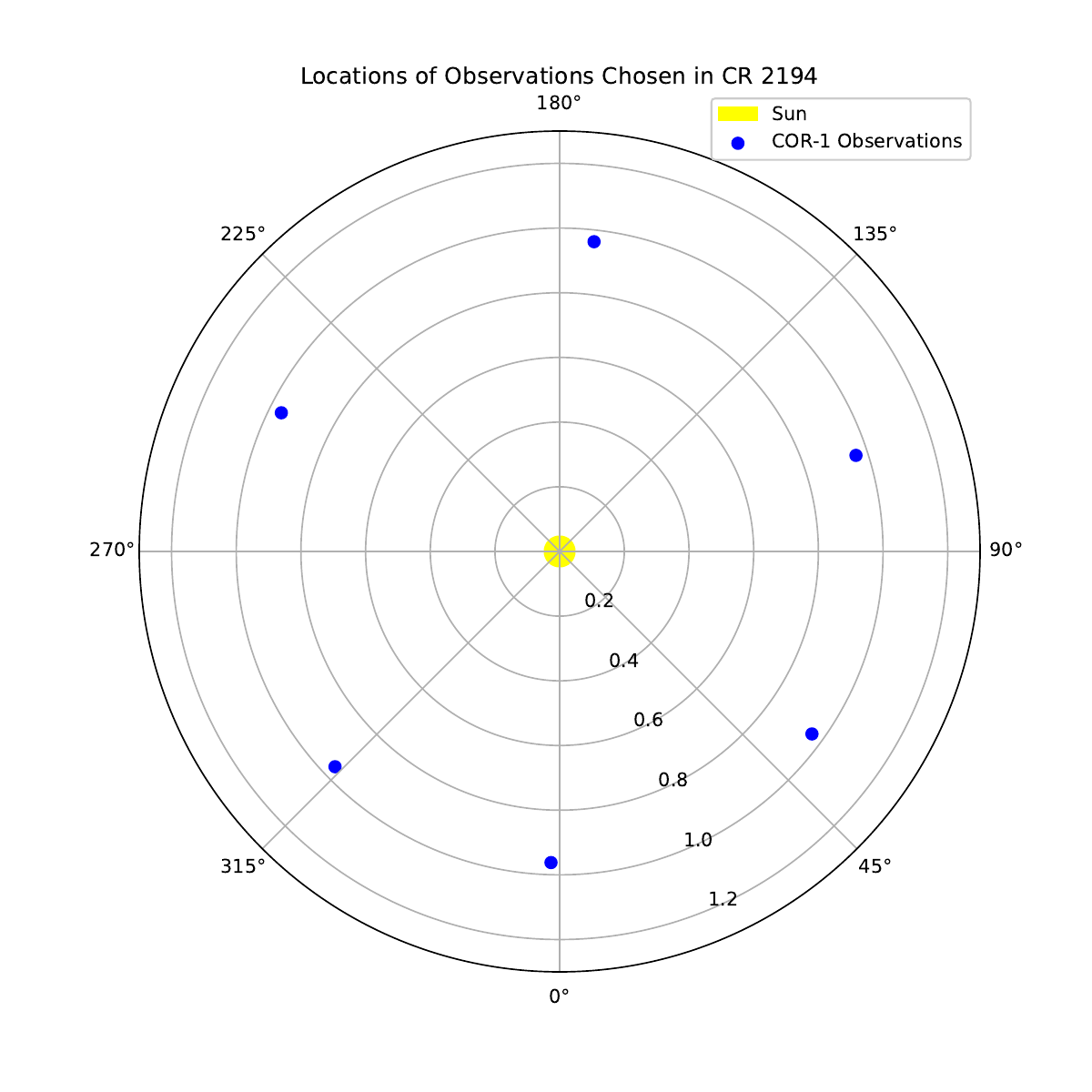}
\centering
\caption{Polar plot showing the Carrington longitudes of each \acrshort{cor-1} observation chosen in this study. These six observations correspond to \acrshort{cr} 2194 and were chosen to provide roughly $60^\circ$ difference of Carrington longitude between each observation.
}
\label{fig:polar_plot}
\end{figure}

To generate a pair of comparison images, we start with the \acrshort{mas} model simulation calculated at the time of the August 2017 solar eclipse.  We use the toolset \citet{helioweb2017heliocentric} to find \acrshort{stereo} A's position in Carrington Heliographic coordinates at a particular time and then query the \acrshort{cor-1} database in order to find observations corresponding to this time. We generate new model images from the \acrshort{mas} solution as described in \textcolor{black}{S}ection \ref{subsec:synthesizing_synthetic_images} from this perspective using the retrieved Carrington Heliographic latitude and longitude. We then output the appropriate model data, (which will be further described in \textcolor{black}{S}ection \ref{subsec:choosing_data_populations}), for use in the segmentation comparison, which will be described in \textcolor{black}{S}ection \ref{subsec:angular_discrepancies}. We then determine the next observation time corresponding to a $60^\circ$ rotation of longitude and start the process again, resulting in a pair of six model slices in an approximately $360 ^{\circ}$ span and accordingly six white-light images for each observatory taken over a 17 day span at the same perspective as the model images in the \acrshort{pos} view. This is done to generate a dataset averaged over a rotation of the Sun in order to have an unbiased comparison with regards \textcolor{black}{its} viewing angle. Figure \ref{fig:polar_plot} shows the longitudinal distribution of these six observations across \acrshort{cr} 2194, which corresponds to the time period of the 2017 solar eclipse and this study.

\textcolor{black}{We generate the \acrshort{mas} \acrshort{pos B} by translating the \acrshort{mas} \acrshort{B} from radial to cartesian coordinates. The \acrshort{mas} \acrshort{B} is generated from a \acrshort{mas} data cube via the \textit{for\_dump} command in \acrshort{forward}, and then the \acrshort{pos} \acrshort{B} is calculated by indexing the central slice of this data cube, which corresponds to the \acrshort{pos}. This data is then saved as a FITS file which allows for coalignment as described in Section \ref{subsec:synthesizing_synthetic_images}.}

\begin{figure}
\noindent\includegraphics[width=\textwidth]{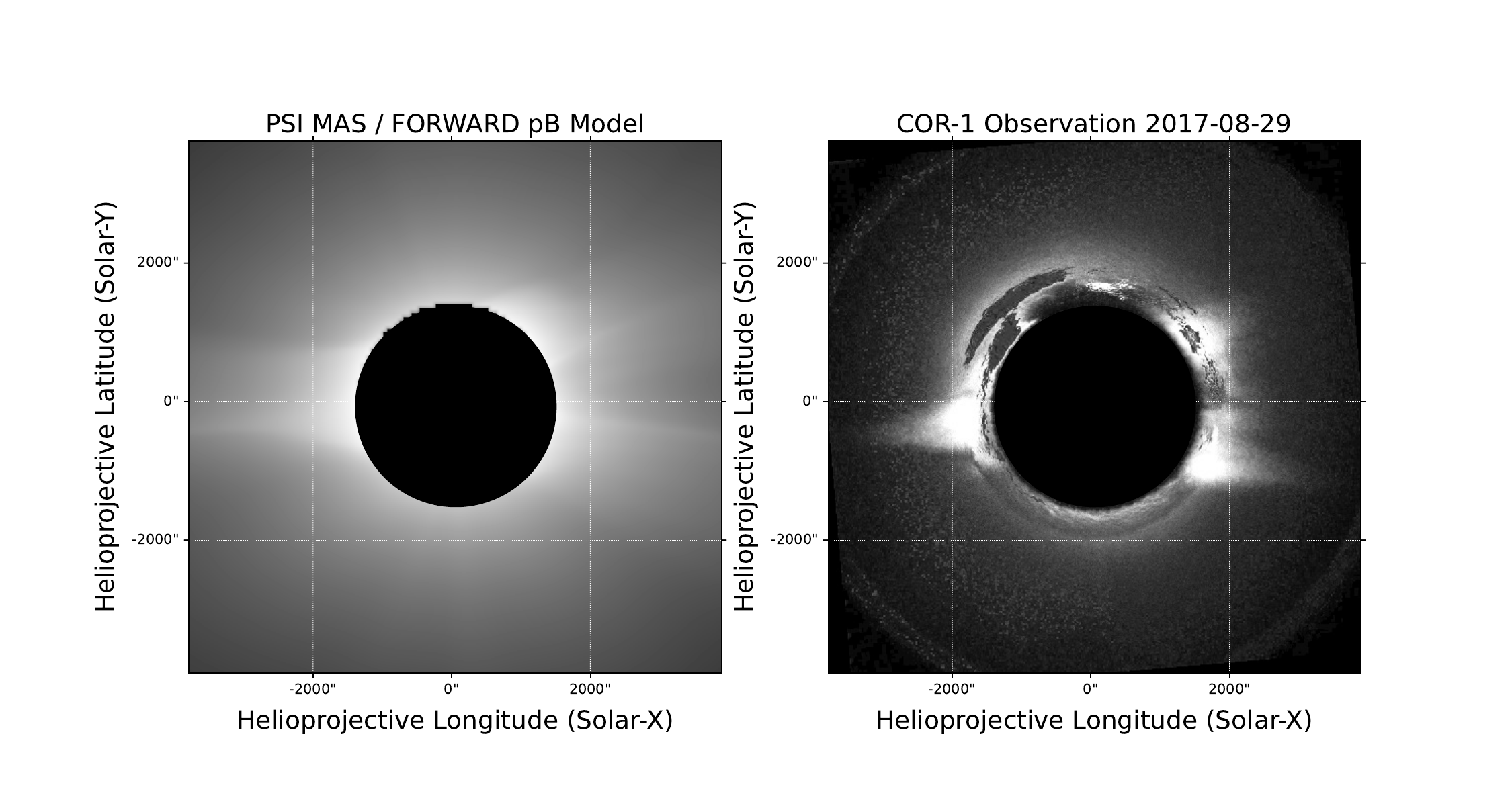}
\caption{(\textcolor{black}{Left}) Synthetic \acrshort{forward pB} image co-aligned to the equivalent observational perspective as the \acrshort{cor-1} observation. \textcolor{black}{(Right)} \acrshort{cor-1 pB} observation on 2017-08-29. }
\label{fig:cor-1_pB_model_comparison}
\end{figure}

\begin{figure}
\begin{center}
\hspace{1.4 cm} PSI MAS $n_e$ Eclipse Model \hspace{4.4 cm} COR-1 Observation 2017-08-29

\noindent\includegraphics[width=\textwidth]{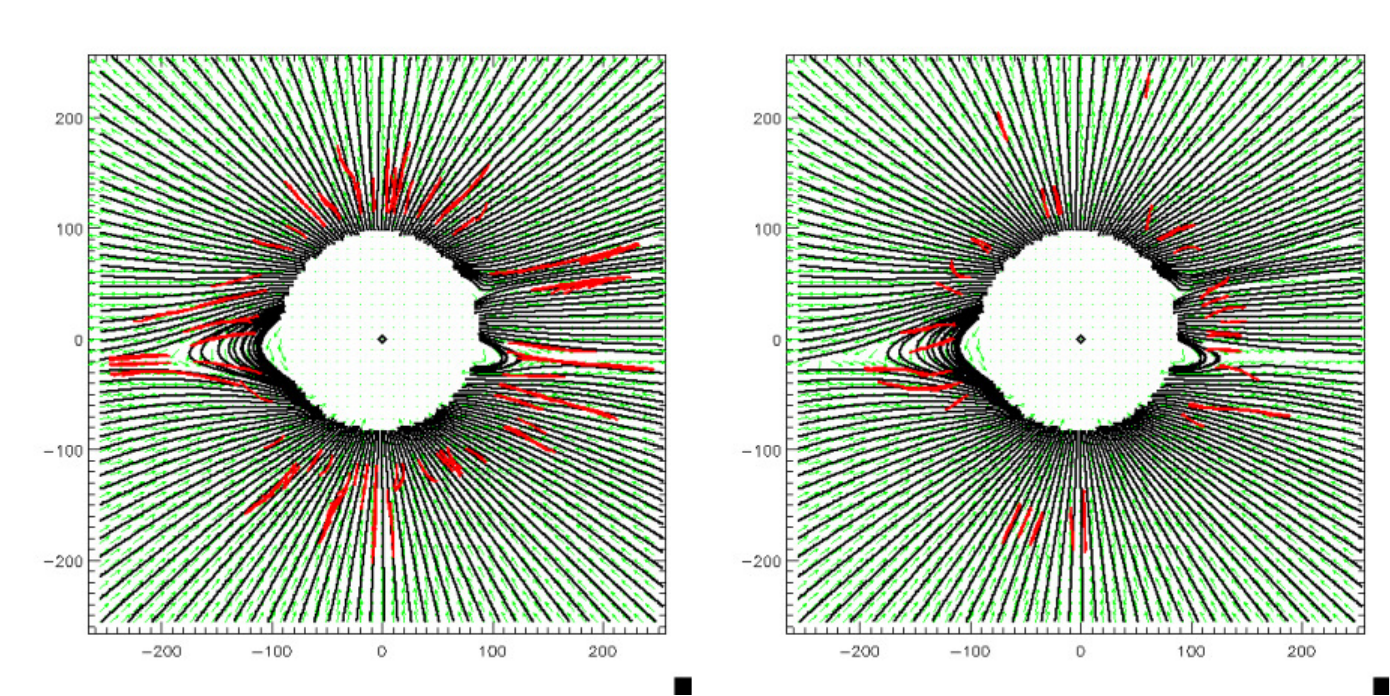}
\caption{(Left) \textcolor{black}{\acrshort{mas}} \acrfull{pos B} orientation with QRaFT filtered features from \acrshort{ne central} model \textcolor{black}{image} for 2017-08-29. (Right) Model \acrshort{pos B} orientation with QRaFT filtered features from \acrshort{cor-1} representative median image for 2017-08-29. \textcolor{black}{The red lines in each plot represent the segmented features from \acrshort{qraft}. The green arrows represent the directions of the \acrshort{mas} \acrshort{pos B} for the 2017-08-29 model \textcolor{black}{image}, while the black lines represent the corresponding magnetic field lines.}}
\label{fig:centralB}
\end{center}
\end{figure}

Figure \ref{fig:cor-1_pB_model_comparison} shows a comparison of a \acrshort{cor-1} observation on 2017-08-29, generated as described in \textcolor{black}{S}ection \ref{subsec:observations}, along with a corresponding \acrshort{forward pB} model \textcolor{black}{image}. \textcolor{black}{Figure \ref{fig:centralB} shows the \acrshort{mas} \acrshort{pos B} orientation, (green arrows representing the direction of the \acrshort{B} and black lines representing the magnetic field lines), with \acrshort{qraft} segmented features overplotted in red for the (left) \acrshort{ne central} and (right) \acrshort{cor-1 pB} for 2017-08-29.}

% Since we co-align these synthetic \acrshort{pB} images to the \acrshort{cor-1} observations respectively, and compare each from an equivalent perspective in space, we are thus directly comparing the orientation of the observed corona to the model generated synthetic corona.

\begin{figure}
\centering
	\noindent\includegraphics[width=0.5\textwidth]{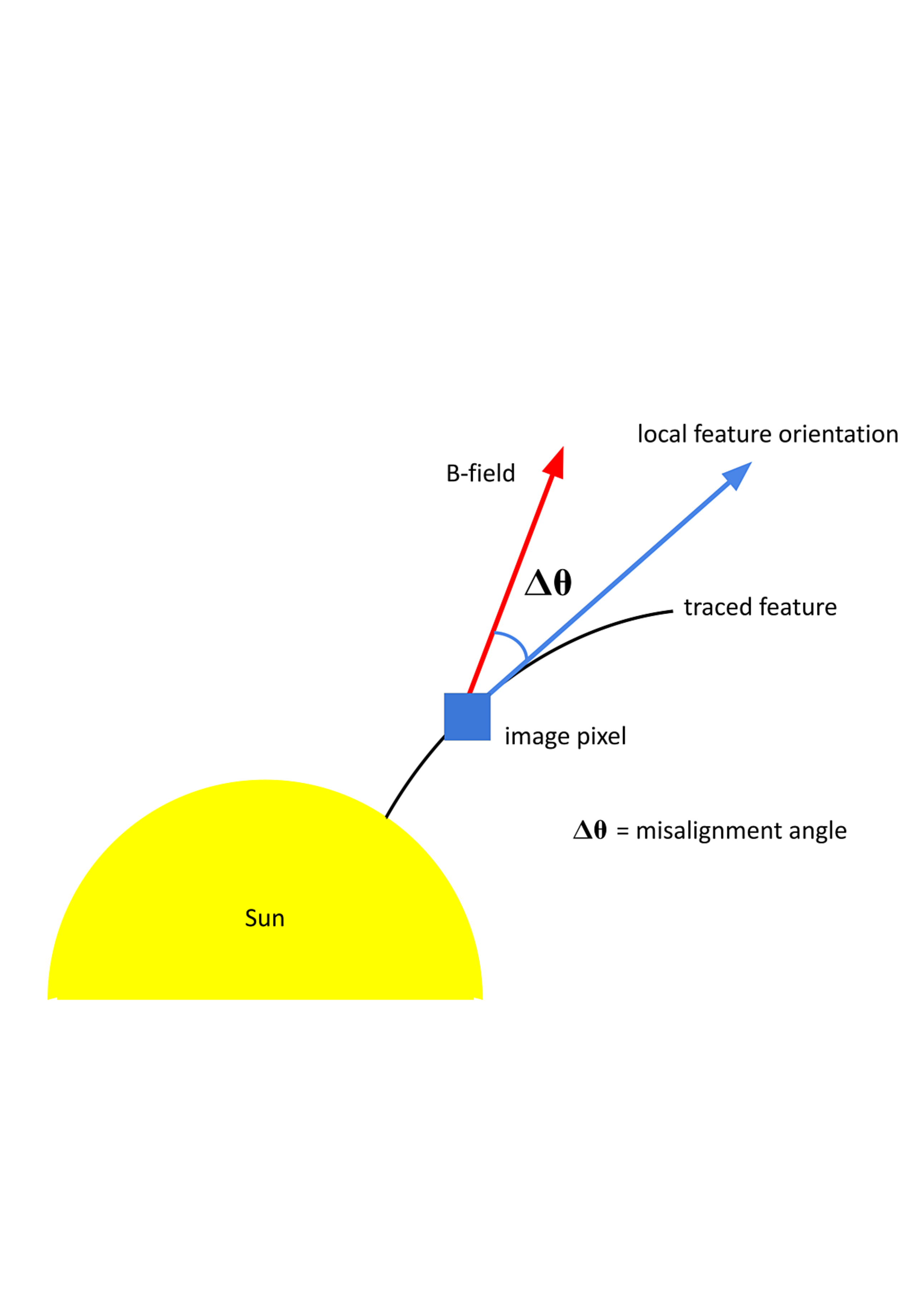}
	\caption{Sketch of \textcolor{black}{\acrshort{angular difference}} calculation used to determine performance of the \acrshort{qraft} feature tracing algorithm. For each pixel the expected magnetic field orientation, given by the PSI model's $B_z$ vs $B_y$ magnetic field components, is compared to the orientation of the segmented feature determined by the \acrshort{qraft} algorithm. The \textcolor{black}{\acrshort{angular difference}} is the statistic which determines how well \textcolor{black}{\acrshort{qraft}} segmented each feature compared to \textcolor{black}{the \acrshort{mas} \acrshort{pos B}}, with a \textcolor{black}{\acrshort{angular difference}} of $0 ^{\circ}$ indicating a perfect fit.
}
	\label{fig:method_sketch}
\end{figure}

\subsection{Calculating \textcolor{black}{\acrfull{angular difference}}}
\label{subsec:angular_discrepancies}

As in \citet{jones2016, jones2020}, we utilize an \textcolor{black}{\acrfull{angular difference}} scheme that compares the orientation of the features segmented by \acrshort{qraft} to the projected orientation of the model magnetic field for both real and synthetic \acrshort{pB} images. This scheme is shown in \textcolor{black}{F}igure \ref{fig:method_sketch} and is the beginning stage of the final component of our framework, (\textcolor{black}{F}igure \ref{fig:framework_overview}). By comparing the \textcolor{black}{\acrshort{angular difference}} results \textcolor{black}{from the real and} synthetic \acrshort{pB} images, we are able to evaluate how well the \acrshort{qraft} algorithm performs. 

% In order to generate a statistic on how well the \acrshort{qraft} feature tracing method determined the orientation of each feature to the orientation of the model field, we find the \textcolor{black}{\acrshort{angular difference}} between the orientation of the traced feature to the orientation of the model field, as shown in \textcolor{black}{F}igure \ref{fig:method_sketch}. 
For each \textcolor{black}{pixel of the} traced feature, we compare the orientation of the slope of the polynomial that approximates the feature's orientation to the expected magnetic orientation from the model. The expected magnetic orientation of the model field is given by the $B_z$ vs $B_y$ magnetic field components, given that we are using Carrington Heliographic coordinates in the \acrlong{pos} view. A \textcolor{black}{\acrshort{angular difference}} of $0^{\circ}$ would constitute a perfect match between the orientation of the traced feature and the expected magnetic orientation from the model at a given pixel. 

We define $\vec{v_1}$ as the orientation of the plasma density feature traced by \acrshort{qraft} in \acrshort{pos} coordinates. We define this orientation as:

\begin{equation}
    \vec{v_1}_{i,k} = \begin{bmatrix} 
x_{k+1} - x_k \\[4pt] 
y_{k+1} - y_k 
\end{bmatrix} \qquad
\end{equation}

where \textcolor{black}{$x_{k+1} - x_k$} and \textcolor{black}{$y_{k+1} - y_k$} are the horizontal and vertical components \textcolor{black}{of the vector between} nodes \textcolor{black}{$k$ and $k+1$} in the \textcolor{black}{$i^{th}$ segmented feature}. 

We define $\vec{v_2}$ as the orientation of the of the \textcolor{black}{\acrshort{mas} \acrshort{pos B}} \textcolor{black}{as described in Section \ref{subsec:data-model-comparison}}. We define th\textcolor{black}{e} 'B-vector' \textcolor{black}{orientation} as:

\begin{equation}
    \vec{v_2}_{i,k} =
\begin{bmatrix} 
B_y(x_k, y_k) \\[4pt] 
B_z(x_k, y_k) 
\end{bmatrix}
\end{equation}

where $B_y$ and $B_z$ are the $y$ and $z$ components of \textcolor{black}{\acrshort{mas} \acrshort{pos B}}. 

We define \textcolor{black}{\acrshort{angular difference}} as the angle between the orientation of \acrshort{qraft}'s polynomial feature, $\vec{v_1}$, and the orientation of the expected magnetic field generated by the model, $\vec{v_2}$, at a given pixel. Since $\vec{v_1} \cdot \vec{v_2} = \left|v_1\right| \left|v_2\right| \cos \theta$ and $ \vec{v_1}^\perp \cdot \vec{v_2} = \left|v_1\right| \left|v_2\right| \sin \theta$ \citep{HILL1994138}, this angle can be calculated using the following formula. 

\begin{equation}
    \Delta \theta =\tan^{-1} \left(\frac{\vec{v_1}^\perp \cdot \vec{v_2}}{\vec{v_1} \cdot \vec{v_2}} \right)  = \tan^{-1} \left( 
\frac{
v_{1x} \ v_{2y} - v_{1y} \ v_{2x}
}{
v_{1x} \ v_{2x} + v_{1y} \ v_{2y}
} \right)
    \label{eq:angular_difference}
\end{equation}

This angle has a range of $-90^\circ <$ \acrshort{angular difference} $< 90^\circ$ and defines the \textcolor{black}{agreement} between the orientation of the predicted model \acrfull{pos B} and the projected orientation of plasma density features given by \acrshort{qraft}. \textcolor{black}{Figure \ref{fig:method_sketch} displays this angle.}

\subsection{Statistical Analysis of \textcolor{black}{\acrlong{angular difference}s}}
\label{subsec:statistical_analysis_methodology}
The final step of our framework, as shown in the second half of component four in \textcolor{black}{F}igure \ref{fig:framework_overview}, is the generation of comprehensive statistics and metrics for this evaluation.
We compute each pixel's \textcolor{black}{\acrshort{angular difference}} for every \textcolor{black}{traced}, and calculate two sets of statistics evaluating \acrshort{qraft}'s \textcolor{black}{performance}. The first set of statistics we calculate are the median and average \textcolor{black}{\acrshort{angular difference}}. We do this to quantify \acrshort{qraft}'s performance, signify outliers\textcolor{black}{,} and compare the similarity of the median and average values of each dataset. Along with the average \textcolor{black}{\acrshort{angular difference}} we also calculate the standard deviation\textcolor{black}{,} $95\%$ confidence interval of \textcolor{black}{\acrshort{angular difference}}\textcolor{black}{, kurtosis, and skewness}. We use these statistics to measure the similarity of \acrshort{qraft}'s performance on white-light coronagraph data when compared to simulated \acrshort{pB} images. \textcolor{black}{Then}, we calculate a histogram and corresponding probability density\textcolor{black}{, (t}he process for calculating a probability density from a Gaussian kernel density estimate is described in appendix \ref{subsec:gaussian_kde}), for each dataset. This process allows us to approximate the discrete distributions of the histograms as a smooth analytical function that can be used to generate statistical comparisons of the dataset.

% The second set of statistics we calculate are more sophisticated metrics aimed at comparing the properties of each dataset to one another. These include the \acrfull{jsd}, kurtosis, skew, and \acrfull{hsd}. 

\textcolor{black}{To evaluate} the similarities of the two distributions, we use the \acrfull{jsd}\textcolor{black}{, and the \acrfull{kld}}\textcolor{black}{, \textcolor{black}{both} defined} in appendix \ref{subsec:JSD_KLD}\textcolor{black}{, which determines} how similar the shapes of two probability distributions are. The closer \textcolor{black}{these} statistic\textcolor{black}{s are} to $0$, the more similar the two distributions are. The probability distributions are normalized by amplitude when evaluating \textcolor{black}{these statistics}.

\textcolor{black}{Finally, we} use \acrfull{hsd}\textcolor{black}{,} described in appendix \ref{subsec:Tukey_HSD}\textcolor{black}{,} to evaluate the relative difference in mean values between each data population, \textcolor{black}{providing a measure of their relative error contributions}. \textcolor{black}{I}n our use of the \acrshort{hsd}, we evaluate this statistic at $\alpha=0.05$, corresponding to a $95\%$ confidence interval. Therefore, in each evaluation, this metric determines if the difference of means between each population is statistically significant corresponding to a $95\%$ confidence interval.

\subsection{Choosing Data Populations in which to Test Framework}

\label{subsec:choosing_data_populations}

In order to test the data to model comparison and decompose the relative error contributions based on our assumptions, we choose four populations when comparing the spatial geometry of features segmented by \acrshort{qraft} to the \textcolor{black}{\acrshort{mas} \acrshort{pos B}}. First, we retain the \acrfull{ne central} of each model \textcolor{black}{image} to use as a one-to-one comparison of the orientation of plasma density features in the \acrlong{pos} to the \textcolor{black}{\acrshort{mas} \acrshort{pos B}}. We then retain the \acrfull{ne los} as the second population. We then retain the \acrfull{forward pB} calculation as the third population for each model \textcolor{black}{image}. \textcolor{black}{F}inally\textcolor{black}{, we} use the corresponding \textcolor{black}{\acrshort{cor-1 pB}} observations as the fourth population. 

These populations are selected in order to provide a varying degree of model complexity when compar\textcolor{black}{ing} to the \textcolor{black}{\acrshort{mas} \acrshort{pos B}}. The \acrshort{ne central} provides a base one to one comparison with the \textcolor{black}{\acrshort{mas} \acrshort{pos B}}. The \acrshort{ne los} adds a degree of complexity by integrating the electron density along the \acrlong{los} of the observer. The \acrshort{forward pB} introduces the geometric scattering \textcolor{black}{effect} along the \acrshort{los} into the integration. Finally, the white-light \textcolor{black}{\acrshort{cor-1 pB}} observations provide the highest degree of complexity as the observational data. The consequences of the differences of resulting statistics between each of these populations will be discussed further in \textcolor{black}{S}ection \ref{subsec:sources_of_error}. It should be noted that these potential consequences lie somewhat outside of the scope of this paper, which is defining and showcasing the actual methodology of this framework, (\textcolor{black}{F}igure \ref{fig:framework_overview}). Nonetheless these consequences hint at potential areas of further study utilizing this framework, which is why we choose to include them in our discussion.

\section{Results}
\label{sec:results}

% \begin{figure}
% 	\noindent\includegraphics[width=\textwidth]{forward_pB-cor1-figure.png}
%     \caption{(Left) \acrshort{ne central} Model slice for 2017-08-20 with resulting \acrshort{qraft} angular error plotted by position of pixel. (Right) \acrshort{cor-1} observation for 2017-08-20 with resulting \acrshort{qraft} angular error plotted by position of pixel. Potential sources of error are discussed in section \ref{subsec:sources_of_error}.}
%     \label{fig:1x2Angle_Err_Example}
% \end{figure}

\begin{figure}
\begin{center}
\includegraphics[width=8.8 cm]{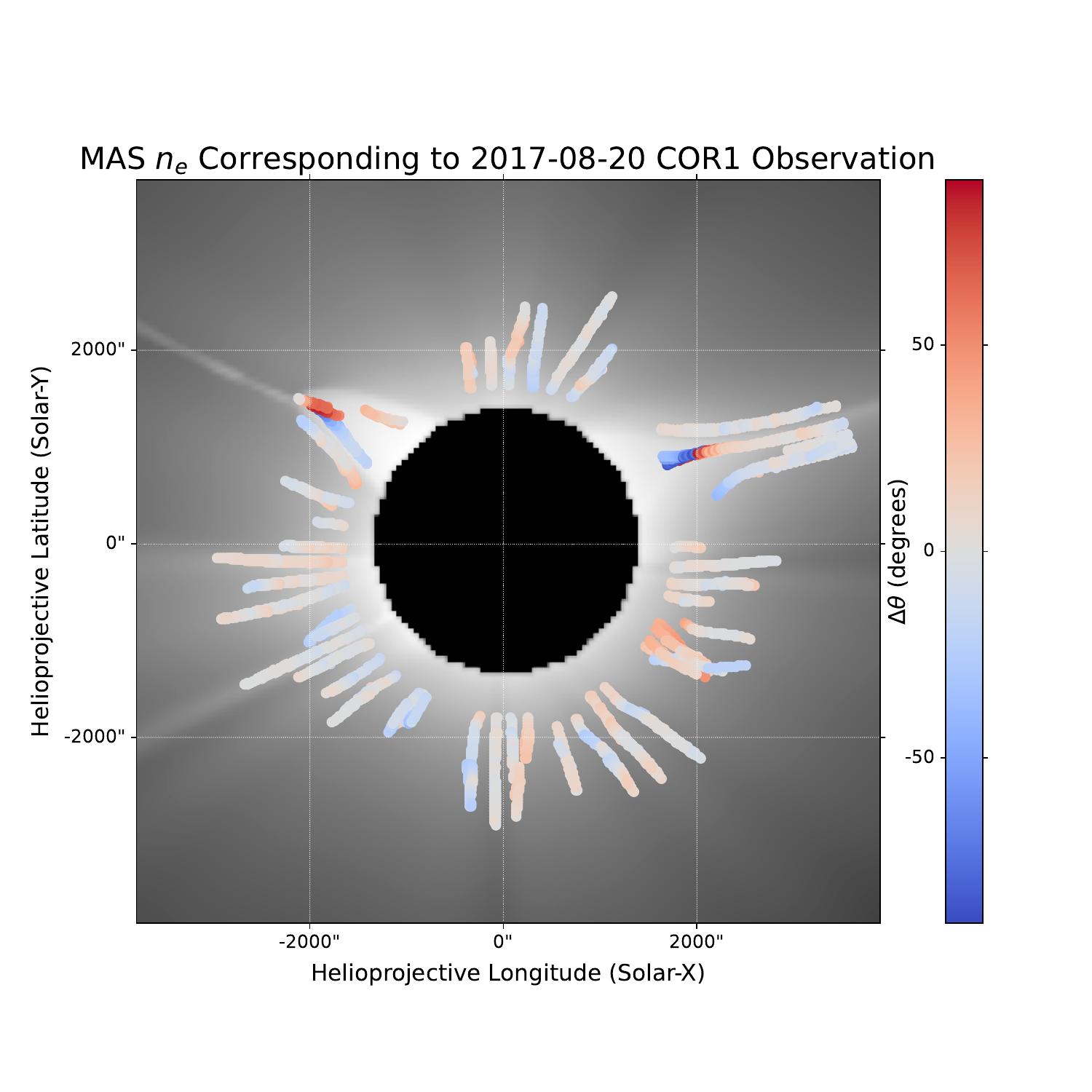} \,\,\,\, \includegraphics[width=8.8 cm]{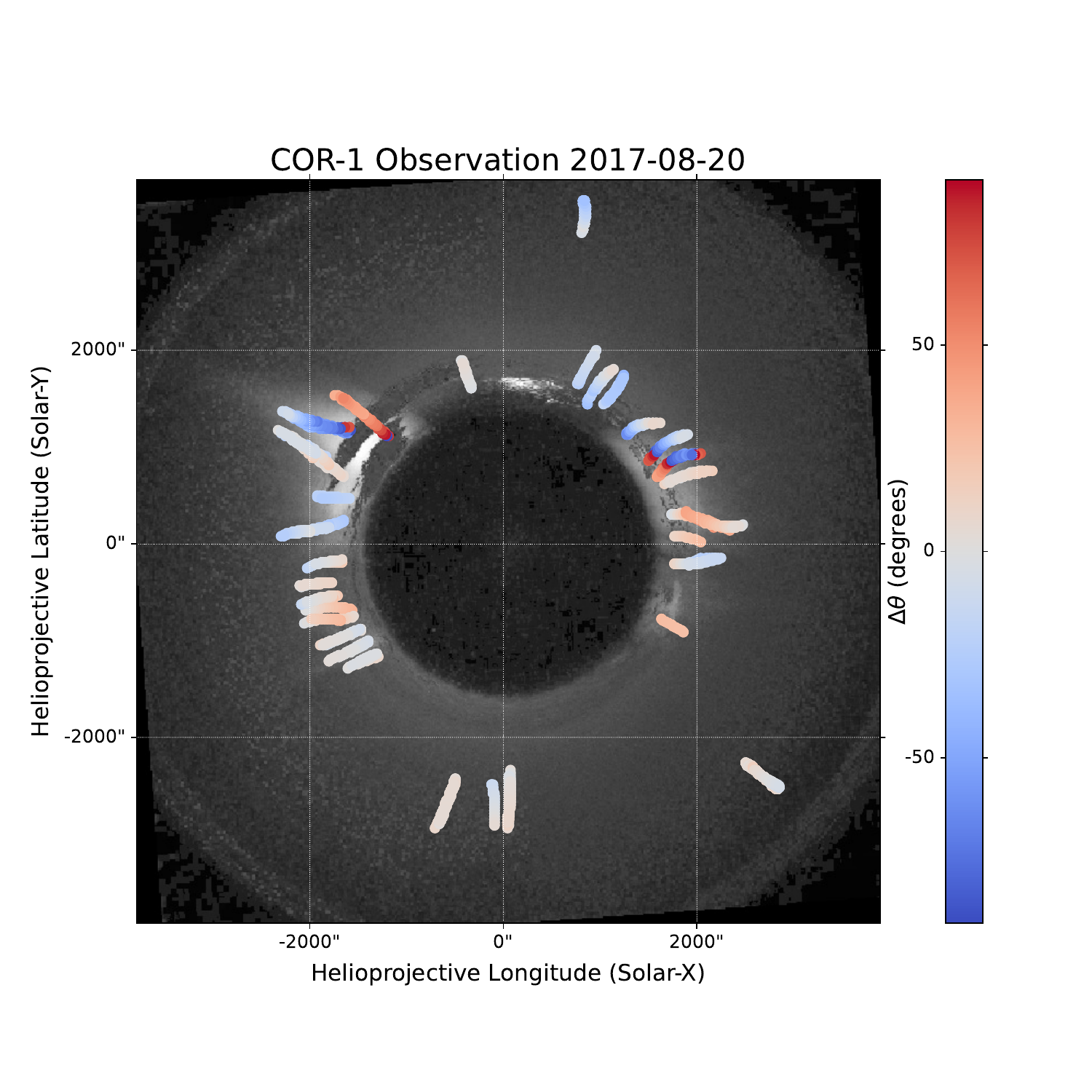}
\end{center}
\caption{(Left) \acrshort{ne central} Model \textcolor{black}{image} \textcolor{black}{corresponding to} 2017-08-20 \textcolor{black}{\acrshort{cor-1 pB} observation} with resulting \textcolor{black}{\acrshort{angular difference}} plotted by position of \textcolor{black}{\acrshort{qraft} feature node. (Right) \acrshort{cor-1 pB}} observation for 2017-08-20 with resulting \textcolor{black}{\acrshort{angular difference}} plotted by position of \textcolor{black}{\acrshort{qraft} feature node}. Potential sources of error are discussed in Section \ref{subsec:sources_of_error}.}
\label{fig:1x2Angle_Err_Example}
\end{figure}

\begin{figure}
	\noindent\includegraphics[width=\textwidth]{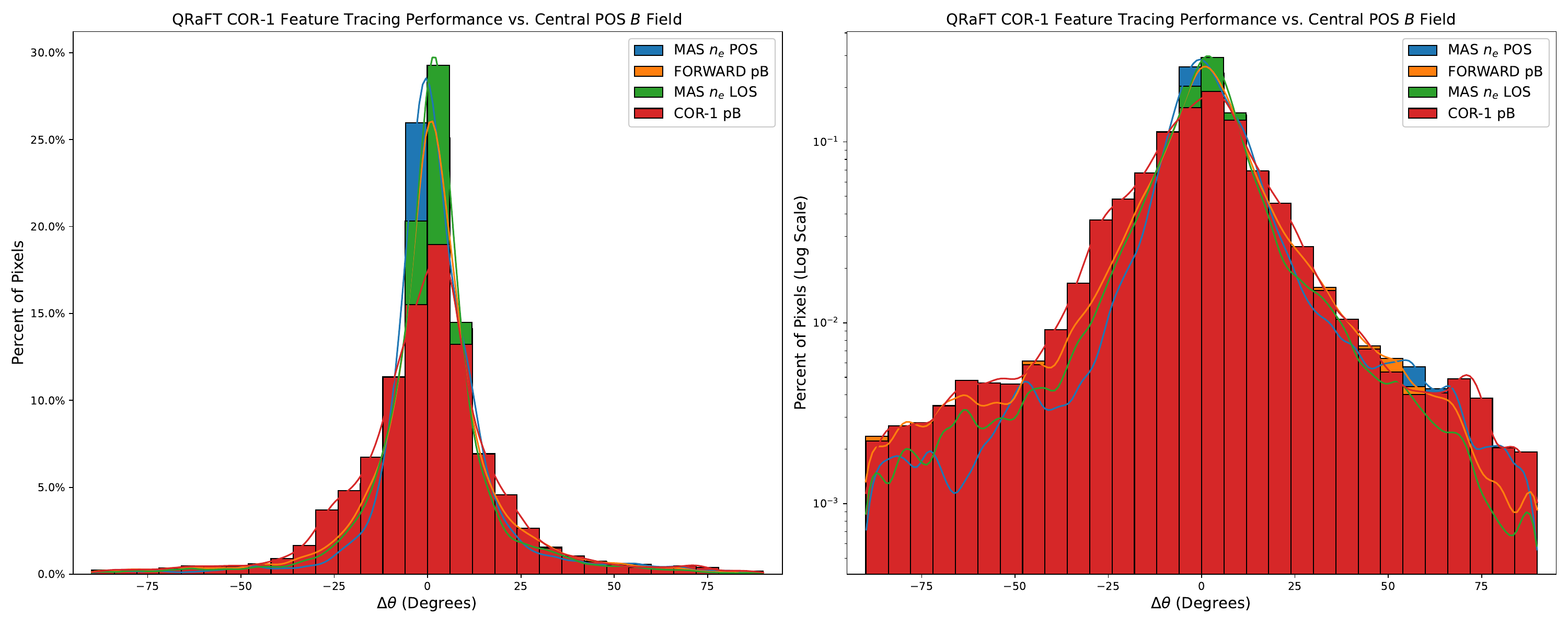}
	\caption{(Left) linear and (Right) log-scaled histograms (with probability densities) of the combined \textcolor{black}{\acrshort{angular difference}} of the \acrshort{qraft} filtered features on the \textcolor{black}{\acrshort{cor-1 pB}} observations and model datasets including \acrshort{ne central}, \acrshort{ne los}, and \acrshort{forward pB} when compared to the expected magnetic orientation from the \acrshort{pos B} from the \acrshort{psi} \acrshort{mas} model. The probability distributions are generated using a kernel density approximation as described in appendix \ref{subsec:gaussian_kde}.}
\label{fig:combined_performance}
\end{figure}

\subsection{Average and Median \textcolor{black}{\acrshort{angular difference}}}

Figure \ref{fig:1x2Angle_Err_Example} shows the \textcolor{black}{\acrshort{angular difference}} \textcolor{black}{for} the \acrshort{qraft} feature tracing method \textcolor{black}{applied} on an \textcolor{black}{image} of the \textcolor{black}{\acrshort{ne central}} and \textcolor{black}{\acrshort{cor-1 pB}} compared to the orientation of the \textcolor{black}{\acrshort{mas} \acrshort{pos B}}. \textcolor{black}{The color scale indicates the magnitude of \acrshort{angular difference} for each of each feature. The distribution of the \acrshort{angular difference} values for each data population are shown in Figure \ref{fig:combined_performance}.} \textcolor{black}{The solid lines} are probability densities corresponding to each histogram generated using a Gaussian kernel density estimation\textcolor{black}{. This process is described in appendix \ref{subsec:gaussian_kde}.} The left plot show these histograms in a linear scale and the right plot show them in a log scale. The behavior of the tails of each of these histograms can be easily examined in the log scale, illuminating which data populations' tails drop off and flatten out relative to each other. \textcolor{black}{Note also} that the \textcolor{black}{peak in the} \textcolor{black}{\acrshort{cor-1 pB}} histogram has a much lower amplitude than the other three histograms, which are all based on data generated from the model. We believe this to be due to \acrshort{qraft} detecting more features, and these features having more pixels, in the model datasets than in the \textcolor{black}{\acrshort{cor-1 pB}} observations. 
 
Table \ref{tab:results_by_date_cor1} shows the average and median values of \textcolor{black}{ the \acrfull{absolute angular difference}} for each data population, as well as the standard deviation, total number of pixels, and $95\%$ confidence interval range, organized by date of observation and corresponding model datasets. In this table we also calculate the kurtosis and skewness\textcolor{black}{, defined in appendix \ref{subsec:kurtosis_skewness},} of the resulting Gaussian kernel density estimate. Table \ref{tab:combined_results_cor1} shows these same statistic\textcolor{black}{s} combin\textcolor{black}{ing all dates for each data population}.

\begin{figure}
\vspace{12pt}
\centering	
 \noindent\includegraphics[scale=0.65]{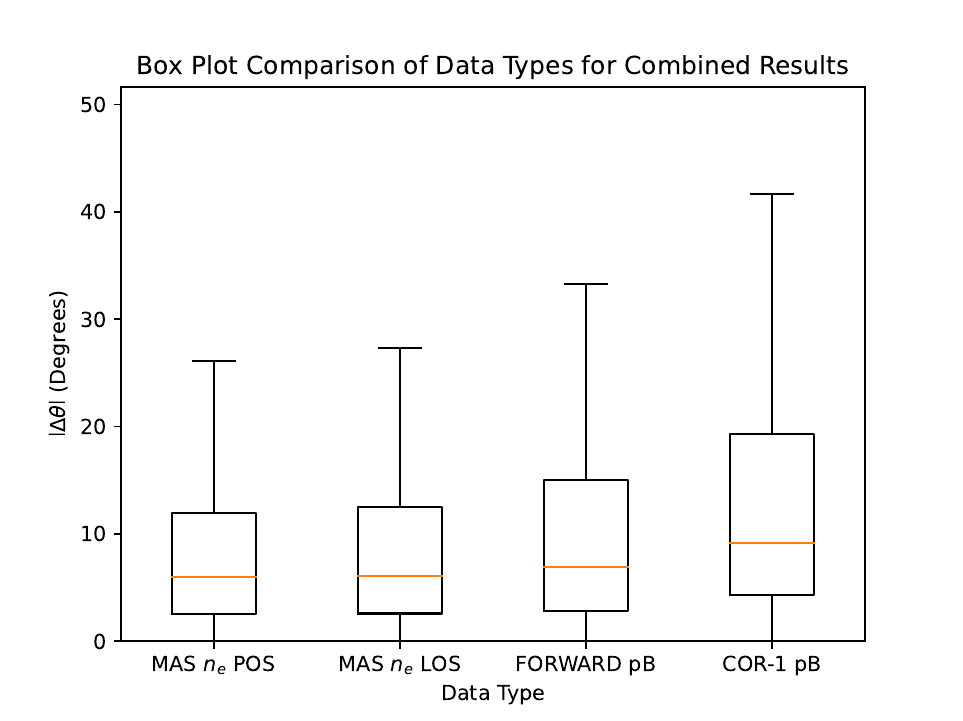}
	\caption{Boxplot of aggregated \textcolor{black}{\textcolor{black}{\acrshort{absolute angular difference}}} for each data population described in \textcolor{black}{S}ection \ref{subsec:choosing_data_populations}. Outliers above the $\acrshort{Q3}+1.5 * \acrshort{IQR}$ range are not shown in the plot. \textcolor{black}{The red line is the median value. The limits of the box identify the \acrshort{Q1} and \acrshort{Q3}, while the whisker indicates the $\acrshort{Q3}+1.5*\acrshort{IQR}$.}}
\label{fig:box_plots}
\end{figure}

Figure \ref{fig:box_plots} showcases box plot statistics from this aggregated data. \textcolor{black}{T}he \acrshort{ne central} population produces the lowest median \textcolor{black}{\textcolor{black}{\acrshort{absolute angular difference}}}, \textcolor{black}{\acrfull{Q1}, \acrfull{IQR}, \acrfull{Q3}}, and $\acrshort{Q3} + 1.5 * \acrshort{IQR}$. Each of these statistics increase \textcolor{black}{for the other} population, with \acrshort{ne los} \textcolor{black}{slightly exceeding} \acrshort{ne central}, \acrshort{forward pB} \textcolor{black}{showing} moderately higher statistics than \acrshort{ne los}, and \acrshort{cor-1 pB} \textcolor{black}{showing} the highest statistics \textcolor{black}{overall}. These statistics can also be seen in \textcolor{black}{T}able \ref{tab:combined_results_cor1}.

% We found that when compared to the \textcolor{black}{\acrshort{mas} \acrshort{pos B}}, the average \textcolor{black}{\acrshort{angular difference}} of features traced in the \acrshort{cor-1 pB} was $14.554 ^{\circ}$, and \textcolor{black}{$12.119^{\circ}$ for features traced} in the synthetic \acrshort{forward pB}. \textcolor{black}{Similarly,} features traced in the \textcolor{black}{\acrshort{ne los} and \acrshort{ne central} had an average \acrshort{angular difference} of} $10.504^{\circ}$ \textcolor{black}{and} $10.479^{\circ}$\textcolor{black}{, respectively}. Median values for each population were The median \textcolor{black}{\acrshort{angular difference}} of each of these populations were $9.157^{\circ}$ \textcolor{black}{(}\textcolor{black}{\acrshort{cor-1 pB}}\textcolor{black}{)}, $6.883^{\circ}$ \textcolor{black}{(}\acrshort{forward pB}\textcolor{black}{)}, $6.035^{\circ}$ \textcolor{black}{(} \acrshort{ne los} \textcolor{black}{)}, and $5.996^{\circ}$ \textcolor{black}{(}\acrshort{ne central}\textcolor{black}{)}.

\textcolor{black}{In Table \ref{tab:combined_results_cor1}, t}he difference between the median and average \textcolor{black}{{\acrshort{absolute angular difference}}} suggests that the dataset is skewed towards lower values \textcolor{black}{with} large \textcolor{black}{{\acrshort{absolute angular difference}} outliers} increas\textcolor{black}{ing} the average \textcolor{black}{\acrshort{absolute angular difference}}. This shows that \textcolor{black}{on average,} \acrshort{qraft}\textcolor{black}{'s} segmented features \textcolor{black}{provides accurate estimates of} the magnetic field \textcolor{black}{orientation} within \textcolor{black}{$10^\circ$ to $15^\circ$} in each data population. \textcolor{black}{The random population's \acrshort{mean angular difference} of $44.882^\circ$ is significantly higher, adding confidence to the relative accuracy of \acrshort{qraft}'s segmented features, while the resulting \acrshort{mean angular difference} of $10.479 ^\circ \pm 0.118^\circ$ for the \acrshort{ne central} population does suggest that there is baseline error in the assumptions and methodology of \acrshort{qraft}}.

\subsection{\acrshort{jsd}\textcolor{black}{, \acrshort{kld}, and \acrshort{hsd}} Statistical Analysis}
The \textcolor{black}{\acrshort{jsd} and \acrshort{kld}} statistic\textcolor{black}{s} allow us to analyze the similarity of two probability distributions' shapes, which is useful for comparing the similarity of each data population. The \acrshort{jsd}, like the \acrfull{kld}, is defined between $0$ and $1$, where a metric closer to $0$ indicates the two distributions are similar in shape, and a metric closer to 1 indicates the two distributions are \textcolor{black}{dissimilar} in shape.  \textcolor{black}{W}e offer both metrics in our results, however we consider \textcolor{black}{the} \acrshort{jsd} as the more important metric in evaluating the similarity of probability distributions as it is a symmetric statistic. \textcolor{black}{As mentioned in Section \ref{subsec:statistical_analysis_methodology}, the \acrshort{hsd} evaluates the relative difference in mean values between each data population to determine if the difference is statistically significant.}

% Since there is no ground truth precedent in which to compare the orientation of segmented coronal features to, as mentioned in section \ref{subsec:Introduction_overview}, we treat the \acrshort{mas} model solution as our known solution in which to base our comparison on. In order to diagnose the statistical trends between the model and data, we require an appropriate comparison framework and statistical metrics in which to quantify this correlation. We use the model synthetic data compared to its own magnetic field orientation as the base comparison, so that the similarity between these statistics will reflect on the correlation between the data and model in regards to the statistical trend of how well quasi-radial features match the expected magnetic field orientation when segmented from coronal data vs model data. 

% \textcolor{black}{Finally, we} use \acrfull{hsd}\textcolor{black}{,} described in appendix \ref{subsec:Tukey_HSD}\textcolor{black}{,} to evaluate the relative difference in mean values between each data population. Performing this analysis allows us to find the relative error contribution between each population. \textcolor{black}{I}n our use of the \acrshort{hsd}, we evaluate this statistic at $\alpha=0.05$, corresponding to a $95\%$ confidence interval. Therefore, in each evaluation, this metric determines if the difference of means between each population is statistically significant corresponding to a $95\%$ confidence interval. 

\textcolor{black}{Table \ref{tab:_hsd_cor1_results_by_date} shows the results of the \acrshort{jsd}, \acrshort{kld}, and \acrshort{hsd} analyses when comparing each data population across each date of this study, while} Table \ref{tab:combined_hsd_cor1} shows the results of the\textcolor{black}{se} analys\textcolor{black}{e}s on the aggregated data \textcolor{black}{for each data population}.

\textcolor{black}{W}e compare this various data populations' results from the analyses done in \textcolor{black}{T}ables \ref{tab:combined_results_cor1} and \ref{tab:combined_hsd_cor1} on a random dataset as a control analysis. This dataset is the size of the mean of means from the four respective datasets described in \textcolor{black}{S}ection \ref{subsec:choosing_data_populations} and is randomized in a range from $-90^{\circ}$ to $90^{\circ}$. \textcolor{black}{The aggregated \acrshort{jsd} metric is lower than $0.02$ for model and observation comparisons while it is above $0.2$ when the comparison includes the probability distribution of the random dataset. This is what we expected the metric to show in this control case.}

\textcolor{black}{Visually inspecting} \textcolor{black}{Figure} \ref{fig:combined_performance} \textcolor{black}{shows} that there is close correlation between the shapes of the \textcolor{black}{\acrshort{angular difference}} \textcolor{black}{probability} distributions \textcolor{black}{for} features traced in the \acrshort{cor-1 pB}, \acrshort{forward pB}, \acrshort{ne los}, and \acrshort{ne central} populations when compared to the model's magnetic field orientation.

\begin{figure}
\noindent\includegraphics[width=\textwidth]{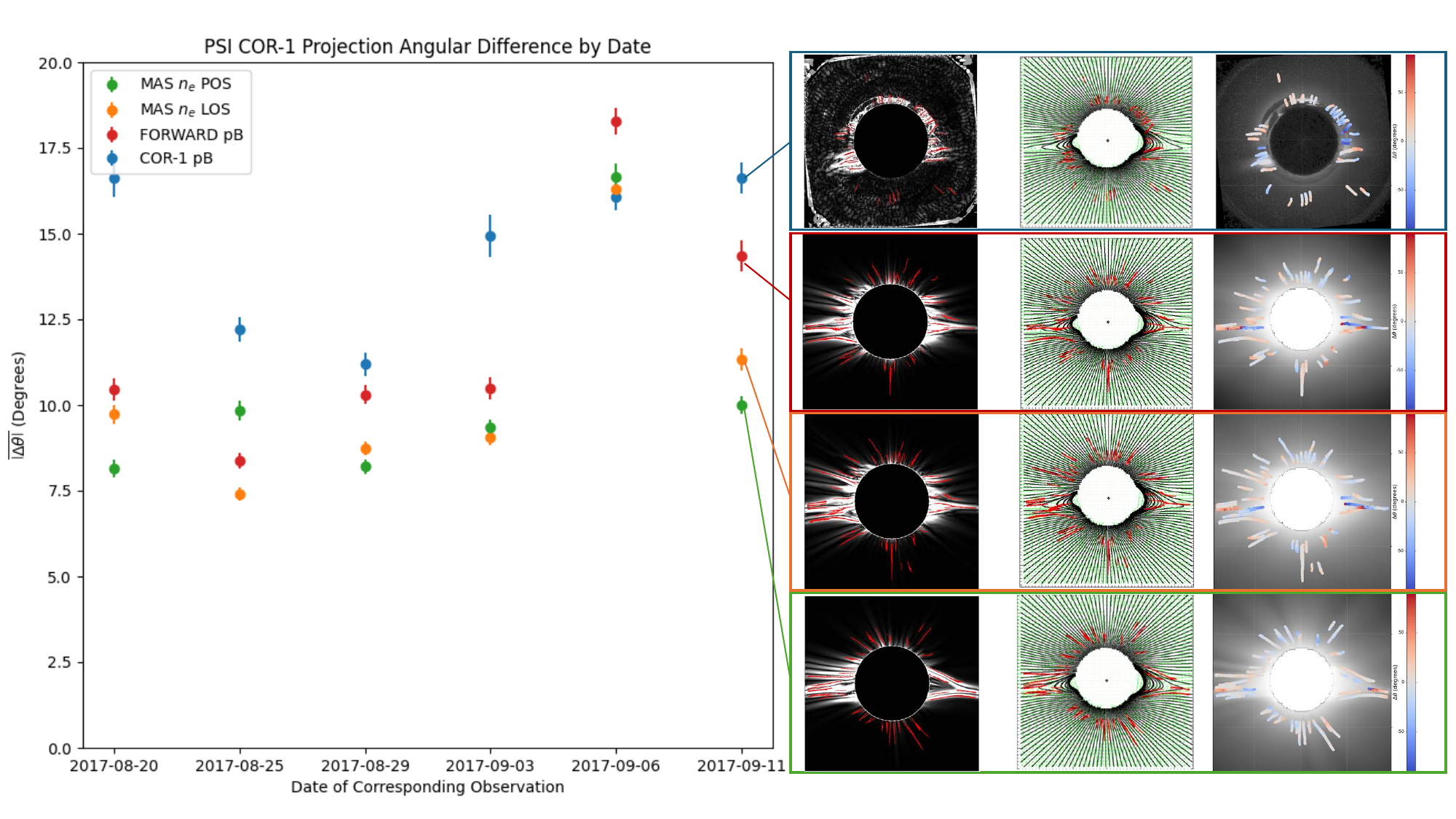}
\caption{Resulting \textcolor{black}{\acrfull{mean angular difference}} by date with examples of each data type shown for 2017-09-11 \acrshort{cor-1} model projection with \textcolor{black}{\acrshort{cor-1 pB}} observation. Error bars represent $95\%$ confidence intervals. These statistics, analyzed by date, can be found in \textcolor{black}{T}able \ref{tab:results_by_date_cor1}. In each set of three example plots for each data point, the left plot shows the \acrshort{qraft} segmented features plotted onto the enhanced image. The middle figure is the \acrshort{qraft} segmented features plotted onto the \acrshort{mas} \acrshort{pos B} orientation as shown in \textcolor{black}{F}igure \ref{fig:centralB}. The right plot is the \textcolor{black}{\acrshort{angular difference}} plotted \textcolor{black}{as a function of \acrshort{qraft} feature node position}, as shown in \textcolor{black}{F}igure \ref{fig:1x2Angle_Err_Example}.}
\label{fig:results_by_date_with_example}
\end{figure}

\begin{table}[ht]
\centering
\caption{Central Tendency Statistics for each population described in \textcolor{black}{S}ection \ref{subsec:choosing_data_populations}. Mean, median, standard deviation, and $95\%$ CI values represent the absolute values in order to display magnitudes. The kurtosis and skewness are calculated on the probability distribution calculated from the signed angle histograms for each population. The table is sorted in ascending order of \textcolor{black}{\acrfull{mean angular difference}} and organized by the date of corresponding \acrshort{cor-1} observation.}
\begin{tabular}{llrrrrrrr}
\hline
 data type             & date       &     \textcolor{black}{\acrshort{mean angular difference}} &   median \textcolor{black}{\acrshort{absolute angular difference}} &      $\sigma_{\Delta \theta}$ &   $95\%$ CI &     n &   kurtosis &   skewness \\
\hline
 \acrshort{ne los}     & 2017-08-25 &  7.408 &    4.477 &  8.949 &       0.184 &  9059 &      8.127 &     -0.003 \\
 \acrshort{ne central} & 2017-08-20 &  8.150 &    4.182 & 12.429 &       0.251 &  9444 &     12.444 &     -0.342 \\
 \acrshort{ne central} & 2017-08-29 &  8.202 &    5.219 & 10.214 &       0.204 &  9584 &      8.884 &      1.028 \\
 \acrshort{forward pB} & 2017-08-25 &  8.368 &    4.656 &  9.997 &       0.227 &  7473 &      4.945 &     -0.103 \\
 \acrshort{ne los}     & 2017-08-29 &  8.736 &    5.294 & 11.454 &       0.191 & 13878 &      9.642 &     -0.157 \\
 \acrshort{ne los}     & 2017-09-03 &  9.062 &    5.802 &  9.724 &       0.236 &  6512 &      3.402 &      0.592 \\
 \acrshort{ne central} & 2017-09-03 &  9.344 &    6.063 & 10.480 &       0.249 &  6793 &      6.902 &     -0.722 \\
 \acrshort{ne los}     & 2017-08-20 &  9.724 &    5.234 & 12.902 &       0.265 &  9077 &      6.654 &      0.505 \\
 \acrshort{ne central} & 2017-08-25 &  9.848 &    5.151 & 14.176 &       0.297 &  8739 &      8.983 &      0.404 \\
 \acrshort{ne central} & 2017-09-11 & 10.008 &    7.179 & 11.774 &       0.252 &  8409 &      8.996 &      0.510 \\
 \acrshort{forward pB} & 2017-08-29 & 10.302 &    5.601 & 14.190 &       0.278 &  9999 &      7.222 &     -0.589 \\
 \acrshort{forward pB} & 2017-08-20 & 10.448 &    5.787 & 13.403 &       0.318 &  6843 &      5.740 &      0.721 \\
 \acrshort{forward pB} & 2017-09-03 & 10.478 &    6.812 & 11.499 &       0.319 &  5001 &      3.365 &      0.503 \\
 \acrshort{cor-1 pB}   & 2017-08-29 & 11.193 &    7.858 & 12.718 &       0.334 &  5568 &      6.179 &      1.712 \\
 \acrshort{ne los}     & 2017-09-11 & 11.333 &    6.577 & 14.279 &       0.324 &  7481 &      5.293 &     -0.657 \\
 \acrshort{cor-1 pB}   & 2017-08-25 & 12.210 &    8.031 & 12.360 &       0.363 &  4455 &      4.446 &      0.117 \\
 \acrshort{forward pB} & 2017-09-11 & 14.350 &    9.227 & 16.636 &       0.456 &  5113 &      4.082 &     -0.461 \\
 \acrshort{cor-1 pB}   & 2017-09-03 & 14.925 &    8.984 & 17.080 &       0.613 &  2981 &      3.855 &      1.692 \\
 \acrshort{cor-1 pB}   & 2017-09-06 & 16.075 &   12.273 & 14.820 &       0.409 &  5047 &      2.141 &     -0.328 \\
 \acrshort{ne los}     & 2017-09-06 & 16.309 &   10.018 & 17.747 &       0.334 & 10821 &      3.165 &     -0.578 \\
 \acrshort{cor-1 pB}   & 2017-08-20 & 16.615 &    8.021 & 19.529 &       0.542 &  4990 &      2.492 &     -0.079 \\
 \acrshort{cor-1 pB}   & 2017-09-11 & 16.633 &   11.681 & 16.153 &       0.454 &  4856 &      1.905 &     -1.312 \\
 \acrshort{ne central} & 2017-09-06 & 16.662 &    9.071 & 18.900 &       0.373 &  9864 &      2.231 &      0.063 \\
 \acrshort{forward pB} & 2017-09-06 & 18.281 &   11.925 & 18.689 &       0.390 &  8836 &      2.548 &     -0.671 \\
\hline
\label{tab:results_by_date_cor1}
\end{tabular}
\end{table}

\begin{table}[ht]
\centering
\caption{Combined central tendency statistics for each population described in \textcolor{black}{S}ection \ref{subsec:choosing_data_populations}. Mean, median, standard deviation, and $95\%$ CI values represent the absolute values in order to display magnitudes. The kurtosis and skewness are calculated on the probability distribution calculated from the signed angle histograms for each population. These populations are compared to a random sample population as a control analysis. The table is sorted in ascending order of \textcolor{black}{\acrshort{mean angular difference}}.}
\begin{tabular}{llrrrrrrr}
\hline
 data type             & date       &     \textcolor{black}{\acrshort{mean angular difference}} &   median \textcolor{black}{\acrshort{absolute angular difference}} &      $\sigma_{\Delta \theta}$ &   $95\%$ CI &     n &   kurtosis &   skewness \\
\hline
  \acrshort{ne central} & combined & 10.479 &    5.996 & 13.855 &       0.118 & 52833 &      7.108 &      0.270 \\
 \acrshort{ne los}     & combined & 10.504 &    6.035 & 13.348 &       0.110 & 56828 &      6.547 &     -0.256 \\
 \acrshort{forward pB} & combined & 12.119 &    6.883 & 14.961 &       0.141 & 43265 &      5.034 &     -0.295 \\
 \acrshort{cor-1 pB}   & combined & 14.554 &    9.157 & 15.694 &       0.184 & 27897 &      3.377 &     -0.003 \\
 random                & combined & \textcolor{black}{44.882} &   \textcolor{black}{45.057} & \textcolor{black}{25.944} &       0.239 & 45205 &     \textcolor{black}{-1.201} &     \textcolor{black}{-0.002} \\
\hline
\label{tab:combined_results_cor1}
\end{tabular}
\end{table}

\section{Discussion}
\label{sec:discussion}

\subsection{Overview}
In this work we presented a quantitative analysis comparing white-light \acrshort{pB} observations to synthetic \acrshort{pB} images from an advanced \acrshort{mhd} model used for predicting the orientation of the solar corona during solar eclipses. Our preliminary results of this framework show close correlation between the performance of quasi-radial coronal feature tracing on both the synthetic and white-light images when compared to the \textcolor{black}{\acrshort{mas}} \textcolor{black}{\acrshort{pos B}}.

The synthetic \acrshort{pB} images generated using \acrshort{mas} establish a ground truth comparison between these \acrshort{forward} modeled \acrshort{pB} images and the underlying 3D electron density and magnetic field, allowing us to test the accuracy of observational methods that approximate the orientation of the coronal \acrlong{B}, such as coronal segmentations. Comparing the orientation of plasma density features extrapolated by \acrshort{qraft} to the \textcolor{black}{model's} expected magnetic orientation establishes a quantifiable measurement of the agreement between each. Having the \acrshort{qraft} algorithm perform similarly on the \textcolor{black}{\acrshort{ne central}}, \textcolor{black}{\acrshort{ne los}}, \textcolor{black}{\acrshort{forward pB}}, and \textcolor{black}{\acrshort{cor-1 pB}} when compared to the expected orientation from the model suggests that the features segmented by \acrshort{qraft} represents coronal magnetic field structure within a reasonable error, and establishes a benchmark to measure the agreement between data and model.

We chose observations that corresponded closely in time with the 2017 solar eclipse, \textcolor{black}{for} which \textcolor{black}{we have a} \textcolor{black}{\acrshort{mas}} \textcolor{black}{simulation}. While \citet{2018NatAs...2..913M} qualitative\textcolor{black}{ly} \textcolor{black}{compared} \acrshort{mas} against eclipse observations, \textcolor{black}{this analysis applied \acrshort{qraft} to quantitatively verify that large-scale quasi-radial features represent true magnetic structure.} These particular features are generally difficult to detect and trace, since the\textcolor{black}{ir} optical signatures are extremely weak \citep{Antonucci2020}. \textcolor{black}{The comparison of model-model and model-observations} allowed us to determine the average and median angular \textcolor{black}{\acrshort{angular difference}} effectively telling us the performance of \acrshort{qraft} in each case. \textcolor{black}{This} is effectively a qualitative analysis on the similarity between the global orientation of the modeled corona when compared to the observed corona, which is similar to how \citet{2018NatAs...2..913M} evaluated their performance in modeling broad aspects of corona. The \acrshort{jsd}, kurtosis, and skew\textcolor{black}{ness} give a \textcolor{black}{quantitative} measurement of how well the \textcolor{black}{probability distributions} represent each other. Minimizing these metrics would then in theory suggest improving the segmentation of quasi-radial features in white-light \acrshort{pB} observations and potentially also improve \acrshort{mhd} modeling as well.

\subsection{Sources of Error}
\label{subsec:sources_of_error}
\textcolor{black}{T}ables \ref{tab:results_by_date_cor1} and \ref{tab:combined_results_cor1} \textcolor{black}{show similar} average and median \textcolor{black}{\acrshort{absolute angular difference} for each population, indicating that segmented features in both the synthetic \acrshort{pB} images and \acrshort{cor-1 pB} observations correlate reasonably well with the \acrshort{mas} \acrshort{pos B}.} \textcolor{black}{The better performance of model based populations with respect} \textcolor{black}{to the \acrshort{cor-1 pB} population} is to be expected \textcolor{black}{as these are model-model comparisons}. Due to the various models, methods, and sources of data used in this analysis, there is ample opportunity for error. We pinpoint four potential causes of error in this analysis.

\subsubsection{Anomalies/artifacts from white-light observations}
\label{subsec:white-light-artifacts-errors}

\textcolor{black}{Several image artifacts are present in the \acrshort{cor-1} observations used in this study. The first artifacts \textcolor{black}{are} most prevalent in the upper left hand portion of the \acrshort{cor-1} observation in \textcolor{black}{F}igures \ref{fig:cor-1_pB_model_comparison} and \ref{fig_qraft_example} \textcolor{black}{due to} too much background subtraction \citep{stereoArtifacts}. The second are \acrshort{cor-1} field lens artifacts \textcolor{black}{that} can be noticed as ring\textcolor{black}{s} around the edge of the solar disk. These artifacts are caused by small defects in the field lens of each \acrshort{cor-1} telescope \citep{stereoArtifacts}.}

\textcolor{black}{Another potential source of error comes from the stacking of \acrshort{cor-1 pB}} images over an 8-hour timespan and \textcolor{black}{calculation of} a median representative image. While this process is effective in enhancing signal and reducing noise, it also adds further potential error, as these structures will smear in the \acrshort{pos} as the Sun rotates. This error can be described mathematically as $\delta \theta = \tan^{-1}\left(\frac{\tan(\lambda)}{\cos(\alpha)}\right) - \lambda$
where $\lambda$ is the solar latitude of a feature, $\alpha$ is the solar rotation angle, and $\delta \theta$ is the angular \acrshort{pos} distortion due to this smearing effect. Over this 8-hour time window, the corona will have rotated $\alpha\sim4.4^\circ$ in longitude. This distortion would be the greatest at a latitude of $\lambda=45^\circ$, resulting in $\delta \theta =0.087^{\circ}$, which would be the maximum contribution to \acrshort{angular difference} from this effect.

% \begin{equation}
%     \delta \theta = \tan^{-1}\left(\frac{\tan(\theta)}{\cos(\alpha)}\right) - \theta
% \end{equation}

During our analysis, we focus on periods of quiet solar activity in order to preserve the large-scale magnetic structure given by the quasi-radial regions. This structure is disrupted if major space weather activity occurs, such as a \acrfull{cme}, since the dynamics of this event causes rapid changes to the orientation of the corona as plasma is ejected outward. There were time periods in which we had to select observations taken at a slightly earlier/later time than predicted \textcolor{black}{because of the presence of a \acrshort{cme} and/or lack of data}. The practical effect this could have on our analysis is that the observations used at the furthest periods from the eclipse may be the least representative of the corona's overall structure in the \acrshort{mhd} simulation, potentially skewing error towards these images.

\subsubsection{Limitations of \acrshort{mhd} models}
\label{subsec:mhd_model_limitations}

Regardless of their level of sophistication, it is important to recognize the sources of error and inherent limitations of global coronal \acrshort{mhd} models. In general, thermodynamic \textcolor{black}{\acrshort{mhd}} models like \textcolor{black}{\acrshort{mas}} require inputs (e.g. magnetic boundary conditions) and parameterization choices (e.g. the coronal heating model), whose sources and parameterizations affect the ensuing 3D distributions of plasma properties and the vector magnetic field.

First, the primary boundary condition used in \acrshort{mas} is a full-sun map of the radial component of the surface magnetic field, $B_r$. This magnetic boundary condition is the most important input to the model, as it largely determines the overall 3D structure and morphology of the coronal magnetic field. Because measurements of the surface magnetic field are typically only available from observatories positioned along the Sun-Earth line (as is the case here), a full-sun map of $B_r$ must be constructed from series of magnetograph observations in time, covering at least one solar rotation (e.g. a synoptic map) or produced by a data assimilative model that also describes the evolution of the field in unobserved region (e.g. synchronic maps produced by surface flux-transport models). This implies a catch 22 for this framework and data validation as a whole, as the  \acrshort{mhd} models must inherently rely on magnetic boundary conditions that are certain to have limitations in both temporal and spatial accuracy. \textcolor{black}{The \acrshort{mas}}  model used a synoptic map from SDO/HMI as the input $B_r$ at the inner boundary to predict the conditions of the corona on August 21, 2017. Since this map was generated using data from \acrshort{cr}s 2192 and 2193, and our observational comparisons use data from \acrshort{cr} 2194, this may be one significant source of error.  This error is challenging to quantify however due to both the observing limitations and the inherent evolution of the Sun's surface magnetic field, which together imply that there is no single map or \acrshort{mhd} model that can be constructed that would simultaneously describe each of our observation dates. That said, the pipeline and error metrics discussed here could in-principle be used to optimize the magnetic boundary conditions for a given case \citep[e.g.;][]{jones2016,jones2020}.

Another important element of \acrshort{mhd} simulations is the combination of the hydrodynamic boundary conditions and coronal heating model that together determine the plasma state of the model (density, temperature, velocity). The heating model in particular is an essential piece of determining the appearance of coronal observables \citep{lionello09}, and there is an intrinsic feedback between the heating, plasma, and the 3D coronal magnetic field that determine where the solar wind is able to open the magnetic field and the morphological appearance of coronal structures \citep[streamers, loops, coronal holes, etc.,][]{downs10}. In other words, this choice may influence the \textcolor{black}{\acrshort{angular difference}} and error metrics discussed here.

Lastly, another potentially relevant element of coronal modeling is the extent to which the presence of large-scale coronal current systems in the form of sheared or twisted filament channels (i.e. free magnetic energy) might affect the observed streamer morphology.  Similarly, the majority of thermodynamic global coronal \acrshort{mhd} models are run as line-tied steady-state relaxations, where the solar wind and heliospheric current sheet(s) are formed but the low coronal domain remains potential to a large extent. There have been recent attempts to use vector magnetic field information and/or magnetofrictional methods to drive global coronal \acrshort{mhd} models to construct coronal current systems that match observations \citep[e.g.][]{hayashi21,hayashi22}, but between the quality limitations of full-sun vector magnetic field observations and the various \textcolor{black}{uncertainties} of boundary driving methods \citep[see][]{tarr24}, this \textcolor{black}{is} extremely challenging.

\subsubsection{Central \acrshort{pos} Assumption}

\label{subsec:central_pos_assumption}

% \textcolor{black}{Another potential source of error is assuming that }

Another potential source of error is \textcolor{black}{that, by using the \acrshort{mas} \acrshort{pos B} as our ground-truth comparison, we assume that the segmented coronal features are located precisely in the \acrshort{pos}.} \textcolor{black}{This assumption is justified by theory,} as \citet{jones2020} notes that \acrshort{pB} images \textcolor{black}{emphasize} coronal material near the \acrlong{pos} \textcolor{black}{due to the scattering geometry of} photospheric light. There are also practical limits justifying this assumption. \textcolor{black}{Comparing the \acrshort{mhd} model's 3D magnetic field requires detailed} \textcolor{black}{3D} topology of the plasma density from \acrshort{pB} observations taken from Earth's perspective. \textcolor{black}{Reconstructing the 3D electron density of the corona is possible using solar tomography, and while recent advances have been made in improving this method using multi-viewpoint data, errors are still present in these reconstructions \citep{2025SoPh..300...46W}. Coronal image segmentation (e.g., \acrshort{qraft}) therefore remains one of the most practical methods to infer the orientation of coronal plasma density structures in the \acrshort{pos}.}

\textcolor{black}{Regardless of the} justifications for this assumption, it \textcolor{black}{introduces possible} error in this analysis. This assumption assumes that the central \acrshort{pos} magnetic field is responsible for the orientation of the large quasi-radial plasma density features that we observe in the corona. Given that the corona is a \textcolor{black}{complex} three-dimensional dynamic structure that varies along the \acrlong{los}, \textcolor{black}{contributions from off-\acrshort{pos} structures to the \acrshort{pB} would distort the alignment with the \acrshort{pos} and introduce error into the comparison with the \acrshort{mas} \acrshort{pos B}, which is inherently planar.} It should be noted from \textcolor{black}{T}able \ref{tab:combined_hsd_cor1} that the difference between the \acrshort{ne central} and \acrshort{ne los} populations was only $0.025^\circ$\textcolor{black}{, and not statistically significant according to the \acrshort{hsd} analysis}, suggest\textcolor{black}{ing} that the error produced from the geometric projection of the 3-D volume of the \acrlong{ne los} onto the \acrshort{pos} was not enough to produce significant error. \textcolor{black}{In contrast, the \acrshort{forward pB} population differed more significantly from the \acrshort{ne los} population, with a mean difference of $1.616^\circ$, and the \acrshort{ne central} population, with a mean difference of $1.641^\circ$. Both of these differences were enough to be statistically significant according to the \acrshort{hsd} analysis. Interestingly, there is an additive nature in these differences, as the sum of the geometric and scattering errors is roughly equivalent to the \acrshort{ne central} - \acrshort{forward pB} comparison. We believe that this is an example of this framework isolating and quantifying distinct error sources that compound on each other in more complex data.} 

The similarity between the \textcolor{black}{\acrshort{forward pB}} and \acrshort{ne los} results suggests that where \acrshort{qraft} is identifying features, the coronal material tends to be near the \acrlong{pos}. \textcolor{black}{This is supported by the small mean difference between these populations. The \acrshort{jsd} distribution shape analysis also supports this, as the \acrshort{jsd} between the \acrshort{ne central} and \acrshort{forward pB} distributions was $0.007^\circ$, and just $0.006^\circ$ between the \acrshort{ne central} and \acrshort{ne los} distributions. Both of these metrics indicate that the shape of each of these probability densities are highly similar, which suggests that while there are measurable mean differences between these populations, their structural similarity remains high.}

\subsubsection{Errors Borne from \acrshort{qraft}}

\label{subsec:qraft_errors}

It is possible, and in some cases likely, that \acrshort{qraft} is not perfectly approximating the orientation of plasma density features in both the \acrshort{cor-1 pB} observations and in the simulated \acrshort{forward pB} images. \textcolor{black}{We briefly overview potential errors contributed from \acrshort{qraft} in this analysis, but \textcolor{black}{they are comprehensively discussed in} \citet{uritsky2025}.} An example of this is \acrshort{qraft}'s 'blob filtering' method not being able to filter out every spurious feature detected by the algorithm. These features are normally small, and when filtered by the radial length are able to be filtered out reasonably well. 

One of the most prominent sources of error stems from \acrshort{qraft} tracing radial features, (or what it perceives to be open field features), in closed field regions of the magnetic field in the \acrshort{pos}. These regions can be especially noticed when plotting the features from \acrshort{qraft} over the expected magnetic field, such as in \textcolor{black}{F}igure \ref{fig:centralB}, and when plotting \textcolor{black}{the} angular error, such as in \textcolor{black}{F}igure \ref{fig:1x2Angle_Err_Example}. These global plots of the angular error allow us to dissect the pattern of how \acrshort{qraft}'s features match the model magnetic field. \textcolor{black}{T}he regions with \textcolor{black}{larger} error correlate to closed regions of the model magnetic field, as the quasi-radial features cross this field near perpendicularly. These errors have a high impact on the resulting statistics of the analysis.

\textcolor{black}{T}he \textcolor{black}{\acrshort{mean angular difference}} of the \acrshort{ne central} population was $10.479^{\circ}$. As this population is a 1-1 comparison of the model \acrshort{pos B} solution to its corresponding \acrshort{ne central}, this figure is a reasonable estimate of the systematic error of \acrshort{qraft}. It is noticeable from \textcolor{black}{F}igure \ref{fig:results_by_date_with_example} that \acrshort{qraft} segments features that do not correspond with true structure in the corona, particularly when comparing segmentations in \textcolor{black}{\acrshort{cor-1 pB}} observations to segmentations in the \textcolor{black}{\acrshort{ne central}}, as shown in \textcolor{black}{F}igure \ref{fig:centralB}. In this comparison, there are clearly spurious segmentations in the \acrshort{cor-1} dataset that do not represent magnetic structure. In particular, features were seen to be originating from locations well beyond the solar occulting radius in several observations. While \textcolor{black}{many of} these features were filtered out \textcolor{black}{via \acrshort{qraft}'s checks} in this analysis, this still suggests that \acrshort{qraft} has room for improvement in feature detection and filtration. Figure \ref{fig_qraft_example} shows steps of the detrending and enhancement method described in \textcolor{black}{S}ection \ref{subsec:qraft}, and in particular panel (d) of this figure shows the fine quasi-radial structure revealed from the second order detrending. There do appear to be artifacts in this enhancement that may be mistaken for features by the algorithm that are not automatically filtered when selecting the final features\textcolor{black}{, especially in the \acrshort{cor-1 pB} example}. Since \acrshort{qraft} is in active development, the conclusion that it can be improved and from this further reduce the baseline error is reasonable.

\subsection{Utilizing Framework to Improve Error}

\begin{table}[ht]
\centering
\caption{Model type comparison using \acrshort{hsd} and \acrshort{jsd} statistics for \acrshort{cor-1} dataset by date. Methods are described in appendices \ref{subsec:Tukey_HSD} and \ref{subsec:JSD_KLD}, respectively. In the (reject \acrshort{H0}?) column, a value of 1 indicates that the \acrfull{H0} is rejected and a value of 0 indicates that \acrshort{H0} is not rejected.}
\begin{tabular}{lllrrrrrrr}
\hline
 group 1                & group 2                & date       &   $\overline{\left|\Delta \theta_2\right|} - \overline{\left|\Delta \theta_1\right|}$ &   lower \acrshort{ci} bound &   upper \acrshort{ci} bound&       \acrshort{kld} &        \acrshort{jsd} &   reject \acrshort{H0}? \\
\hline
 \acrshort{cor-1 pB}   & \acrshort{ne central} & 2017-08-20 &      -8.465 &           -9.102 &           -7.827 & 0.179 & 0.048 &        1 \\
 \acrshort{cor-1 pB}   & \acrshort{ne los}     & 2017-08-20 &      -6.891 &           -7.533 &           -6.249 & 0.117 & 0.032 &        1 \\
 \acrshort{cor-1 pB}   & \acrshort{forward pB} & 2017-08-20 &      -6.167 &           -6.845 &           -5.489 & 0.092 & 0.026 &        1 \\
 \acrshort{ne central} & \acrshort{ne los}     & 2017-08-20 &       1.574 &            1.038 &            2.109 & 0.035 & 0.009 &        1 \\
 \acrshort{ne central} & \acrshort{forward pB} & 2017-08-20 &       2.297 &            1.719 &            2.876 & 0.118 & 0.027 &        1 \\
 \acrshort{ne los}     & \acrshort{forward pB} & 2017-08-20 &       0.724 &            0.141 &            1.307 & 0.060 & 0.014 &        1 \\
 \acrshort{cor-1 pB}   & \acrshort{ne central} & 2017-08-25 &      -2.363 &           -2.905 &           -1.820 & 0.225 & 0.058 &        1 \\
 \acrshort{cor-1 pB}   & \acrshort{ne los}     & 2017-08-25 &      -4.802 &           -5.341 &           -4.263 & 0.110 & 0.030 &        1 \\
 \acrshort{cor-1 pB}   & \acrshort{forward pB} & 2017-08-25 &      -3.842 &           -4.400 &           -3.284 & 0.082 & 0.022 &        1 \\
 \acrshort{ne central} & \acrshort{ne los}     & 2017-08-25 &      -2.439 &           -2.881 &           -1.998 & 0.196 & 0.043 &        1 \\
 \acrshort{ne central} & \acrshort{forward pB} & 2017-08-25 &      -1.479 &           -1.943 &           -1.015 & 0.239 & 0.056 &        1 \\
 \acrshort{ne los}     & \acrshort{forward pB} & 2017-08-25 &       0.960 &            0.500 &            1.421 & 0.022 & 0.005 &        1 \\
 \acrshort{cor-1 pB}   & \acrshort{ne central} & 2017-08-29 &      -2.991 &           -3.516 &           -2.467 & 1.860 & 0.400 &        1 \\
 \acrshort{cor-1 pB}   & \acrshort{ne los}     & 2017-08-29 &      -2.457 &           -2.951 &           -1.963 & 1.826 & 0.365 &        1 \\
 \acrshort{cor-1 pB}   & \acrshort{forward pB} & 2017-08-29 &      -0.891 &           -1.412 &           -0.370 & 1.719 & 0.348 &        1 \\
 \acrshort{ne central} & \acrshort{ne los}     & 2017-08-29 &       0.535 &            0.121 &            0.948 & 0.049 & 0.013 &        1 \\
 \acrshort{ne central} & \acrshort{forward pB} & 2017-08-29 &       2.101 &            1.655 &            2.546 & 0.098 & 0.016 &        1 \\
 \acrshort{ne los}     & \acrshort{forward pB} & 2017-08-29 &       1.566 &            1.157 &            1.974 & 0.016 & 0.004 &        1 \\
 \acrshort{cor-1 pB}   & \acrshort{ne central} & 2017-09-03 &      -5.582 &           -6.240 &           -4.923 & 0.269 & 0.070 &        1 \\
 \acrshort{cor-1 pB}   & \acrshort{ne los}     & 2017-09-03 &      -5.863 &           -6.525 &           -5.200 & 0.352 & 0.084 &        1 \\
 \acrshort{cor-1 pB}   & \acrshort{forward pB} & 2017-09-03 &      -4.447 &           -5.141 &           -3.754 & 0.217 & 0.054 &        1 \\
 \acrshort{ne central} & \acrshort{ne los}     & 2017-09-03 &      -0.281 &           -0.801 &            0.238 & 0.113 & 0.029 &        0 \\
 \acrshort{ne central} & \acrshort{forward pB} & 2017-09-03 &       1.134 &            0.576 &            1.692 & 0.097 & 0.017 &        1 \\
 \acrshort{ne los}     & \acrshort{forward pB} & 2017-09-03 &       1.415 &            0.852 &            1.978 & 0.048 & 0.012 &        1 \\
 \acrshort{cor-1 pB}   & \acrshort{ne central} & 2017-09-06 &       0.587 &           -0.211 &            1.385 & 0.166 & 0.035 &        0 \\
 \acrshort{cor-1 pB}   & \acrshort{ne los}     & 2017-09-06 &       0.235 &           -0.551 &            1.020 & 0.106 & 0.021 &        0 \\
 \acrshort{cor-1 pB}   & \acrshort{forward pB} & 2017-09-06 &       2.206 &            1.393 &            3.019 & 0.103 & 0.019 &        1 \\
 \acrshort{ne central} & \acrshort{ne los}     & 2017-09-06 &      -0.353 &           -0.994 &            0.289 & 0.058 & 0.014 &        0 \\
 \acrshort{ne central} & \acrshort{forward pB} & 2017-09-06 &       1.619 &            0.944 &            2.294 & 0.075 & 0.018 &        1 \\
 \acrshort{ne los}     & \acrshort{forward pB} & 2017-09-06 &       1.972 &            1.310 &            2.632 & 0.017 & 0.004 &        1 \\
 \acrshort{cor-1 pB}   & \acrshort{ne central} & 2017-09-11 &      -6.625 &           -7.292 &           -5.957 & 0.290 & 0.070 &        1 \\
 \acrshort{cor-1 pB}   & \acrshort{ne los}     & 2017-09-11 &      -5.300 &           -5.982 &           -4.617 & 0.219 & 0.051 &        1 \\
 \acrshort{cor-1 pB}   & \acrshort{forward pB} & 2017-09-11 &      -2.283 &           -3.025 &           -1.540 & 0.169 & 0.033 &        1 \\
 \acrshort{ne central} & \acrshort{ne los}     & 2017-09-11 &       1.325 &            0.737 &            1.914 & 0.052 & 0.015 &        1 \\
 \acrshort{ne central} & \acrshort{forward pB} & 2017-09-11 &       4.342 &            3.686 &            4.999 & 0.129 & 0.022 &        1 \\
 \acrshort{ne los}     & \acrshort{forward pB} & 2017-09-11 &       3.017 &            2.345 &            3.689 & 0.040 & 0.010 &        1 \\
\hline
\label{tab:_hsd_cor1_results_by_date}
\end{tabular}
\end{table}

\begin{table}[ht]
\centering
\caption{Model type comparison using \acrshort{hsd} and \acrshort{jsd} statistics for \acrshort{cor-1} combined dataset. Methods are described in sections \ref{subsec:Tukey_HSD} and \ref{subsec:JSD_KLD}, respectively. In the (reject \acrshort{H0}?) column, a value of 1 indicates that the \acrfull{H0} is rejected and a value of 0 indicates that \acrshort{H0} is not rejected.}
\begin{tabular}{lllrrrrrrr}
\hline
 group 1                & group 2                & date     & $\overline{\left|\Delta \theta_2\right|} - \overline{\left|\Delta \theta_1\right|}$ &   lower bound \acrshort{ci} &   upper bound \acrshort{ci} &        \acrshort{kld} &        \acrshort{jsd} &   reject \acrshort{H0}? \\
\hline
 \acrshort{cor-1 pB}   & \acrshort{ne central} & combined &      -4.076 &           -4.424 &           -3.728 & 0.079 & 0.020 &        1 \\
 \acrshort{cor-1 pB}   & \acrshort{ne los}     & combined &      -4.051 &           -4.394 &           -3.707 & 0.056 & 0.014 &        1 \\
 \acrshort{cor-1 pB}   & \acrshort{forward pB} & combined &      -2.435 &           -2.796 &           -2.074 & 0.033 & 0.007 &        1 \\
 \acrshort{cor-1 pB}   & random                & combined &      \textcolor{black}{30.327} &           \textcolor{black}{29.969} &           \textcolor{black}{30.686} & \textcolor{black}{0.885} & \textcolor{black}{0.216} &        1 \\
 \acrshort{ne central} & \acrshort{ne los}     & combined &       0.025 &           -0.259 &            0.309 & 0.022 & 0.006 &        0 \\
 \acrshort{ne central} & \acrshort{forward pB} & combined &       1.641 &            1.336 &            1.946 & 0.026 & 0.007 &        1 \\
 \acrshort{ne central} & random                & combined &      \textcolor{black}{34.403} &           \textcolor{black}{34.102} &           \textcolor{black}{34.705} & \textcolor{black}{1.172} & \textcolor{black}{0.269} &        1 \\
 \acrshort{ne los}     & \acrshort{forward pB} & combined &       1.616 &            1.316 &            1.916 & 0.010 & 0.003 &        1 \\
 \acrshort{ne los}     & random                & combined &      \textcolor{black}{34.378} &           \textcolor{black}{34.081} &           \textcolor{black}{34.675} & \textcolor{black}{1.151} & \textcolor{black}{0.264} &        1 \\
 \acrshort{forward pB} & random                & combined &      \textcolor{black}{32.762} &           \textcolor{black}{32.446} &           \textcolor{black}{33.079} & \textcolor{black}{1.000} & \textcolor{black}{0.232} &        1 \\
\hline
\label{tab:combined_hsd_cor1}
\end{tabular}
\end{table}

% \begin{figure}
% \noindent\includegraphics[width=\textwidth]{COR1_Combined_JSD_no_random_heatmap2.pdf}
% \caption{Heatmap of \acrshort{jsd} statistics of aggregated data populations described in section \ref{subsec:choosing_data_populations}.}
% \label{fig:cor1_combined_jsd_heatmap}
% \end{figure}

\begin{figure}
\begin{center}
\includegraphics[width=8.8 cm]{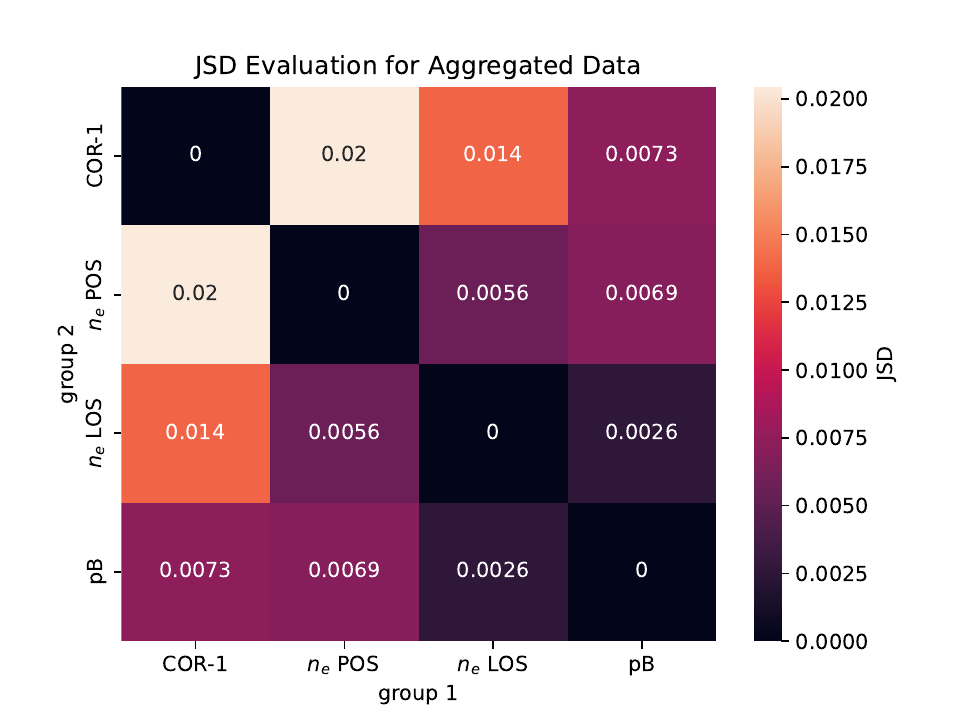} \,\,\,\, \includegraphics[width=8.8 cm]{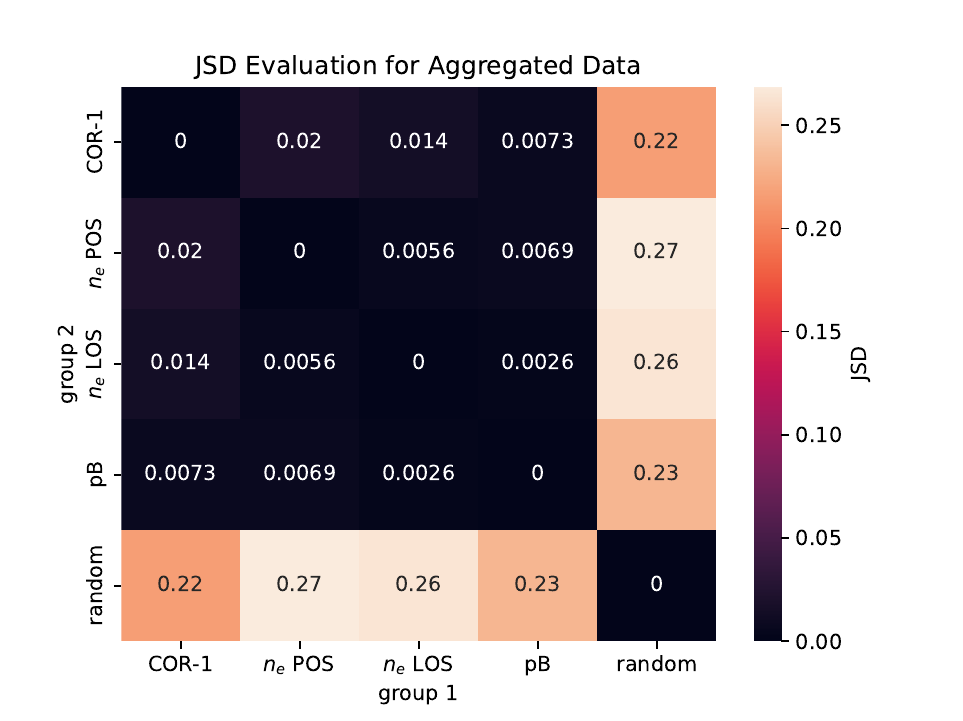}
\caption{Heatmaps of \acrshort{jsd} statistics shown in \textcolor{black}{T}able \ref{tab:combined_hsd_cor1}. The left plot displays a comparison of the \acrshort{jsd} metrics between \acrshort{cor-1 pB}, \acrshort{ne central}, \acrshort{ne los}, and \acrshort{forward pB}. The right plot displays this same comparison, but also includes the random population in the heatmap. In each plot, each diagonal cell has a value of $0$ as by definition the \acrshort{jsd} between two identical probability densities is $0$.}
\label{fig:cor1_combined_jsd_heatmap}
\end{center}
\end{figure}

Through a rigorous set of statistical tests, we believe that we were able to \textcolor{black}{quantitatively} characterize error contributions from these assumed sources. Four \textcolor{black}{data} populations \textcolor{black}{were chosen} to measure this potential error\textcolor{black}{:} \textcolor{black}{\acrshort{ne central}}, \textcolor{black}{\acrshort{ne los}}, \textcolor{black}{\acrshort{forward pB}}, and the \textcolor{black}{observed} white-light \textcolor{black}{\acrshort{cor-1 pB}}. \textcolor{black}{By comparing the \acrfull{mean angular difference} between each population, we can isolate these assumed sources of error. These can be broken down as followed: the geometric projection of the 3-D \acrlong{ne} volume onto the \acrshort{pos} (\acrshort{ne central} vs \acrshort{ne los}), the artificial scattering and other methods used in \acrshort{forward} (\acrshort{ne los} vs \acrshort{forward pB}\textcolor{black}{)}, and all other error contributions associated from the comparison of the modeled \acrshort{pos B} to the segmented features from the observed \acrlong{pB} (\acrshort{forward pB} vs \acrshort{cor-1 pB}). Tables \ref{tab:_hsd_cor1_results_by_date} and \ref{tab:combined_hsd_cor1} summarize these results by date of observation and in aggregate, respectively.}

% The difference between the mean  for these first two populations of the \acrshort{ne central} and the \acrshort{ne los} should provide the error contribution from the geometric projection of the 3-D \acrlong{ne} volume onto the \acrshort{pos}. Likewise the difference between the means of the next two populations given by the \acrshort{ne central} and model calculated \acrshort{forward pB} should provide the error contribution from the artificial scattering and other methods used in \acrshort{forward}. Finally the difference between the last two populations given by the model calculated \acrshort{forward pB} and white-light \acrshort{cor-1 pB} should provide all other error contributions associated from the comparison of the modeled \acrshort{pos B} to the segmented features from the observed \acrlong{pB}. 

% Table \ref{tab:_hsd_cor1_results_by_date} shows the results of this error analysis for each population comparison across each date of observation chosen while table \ref{tab:combined_hsd_cor1} shows the results of this error analysis for each population for the combined data.

While each source of error adds uncertainty, \textcolor{black}{this framework} provides \textcolor{black}{the opportunity} to \textcolor{black}{iteratively} improve \textcolor{black}{accuracy and reduce error}. 
\textcolor{black}{We introduce novel statistics to quantitatively assess the accuracy of coronal feature tracing algorithms in identifying magnetic structure and for how well \acrshort{mhd} simulations represent a specific aspect of observational data. This provides a benchmark for improving both segmentation algorithms such as \acrshort{qraft} and \acrshort{mhd} modeling.} For example, \textcolor{black}{if modifying the \acrshort{mhd} model's} assumptions \textcolor{black}{(e.g., input photospheric magnetic field maps)}  produced better metrics while using this framework, this would suggest a \textcolor{black}{more accurate} representation of the observational data\textcolor{black}{, and thus the observed corona}. Likewise, if a modification to \acrshort{qraft}'s code resulted in better metrics, then this would suggest that \textcolor{black}{it} improved \acrshort{qraft}'s performance and ability to \textcolor{black}{segment} plasma density features. This may unlock insights necessary to improve both \acrshort{mhd} modeling and segmentation algorithms that would not have been obvious by utilizing only qualitative comparisons.

The major advantage of our framework is \textcolor{black}{that complex model-data comparisons can be described in interpretable,} quantitative metrics. \textcolor{black}{An example of this can be seen in Figure \ref{fig:cor1_combined_jsd_heatmap}, which displays heatmaps showcasing the aggregated \acrshort{jsd} metrics with and without a random population included. These heatmaps are effective in showing that comparisons between the most and least realistic datasets yield the highest \acrshort{jsd} values when compared to each other, while comparisons between more similar datasets result in lower metrics. These result\textcolor{black}{s} align with our expectations, reflecting how the metric responds to converging error patterns.}

\textcolor{black}{While this framework could allow us to improve both feature tracing methods and \acrshort{mhd} modeling, it also could compare different varieties of each.} Importantly, this framework has the potential to provide important insight to validating segmentations from white-light data against ground-truth \acrshort{mhd} models, by quantitatively capturing how representative an \acrshort{mhd} model is to the large scale coronal features observed in white-light data. This is important, as ultimately the accuracy of segmentations of large scale plasma density features in the corona directly impacts the accuracy and uncertainty of boundary conditions in space weather models. \textcolor{black}{As an example, the input for \acrshort{wsa} is a full-rotation synoptic map, and inner boundary is a \acrshort{pfss} model, while the outer layer is a ``Schatten" current sheet model used to predict solar wind propagation based on magnetic divergence and coronal hole proximity \citep{Arge2000}. \citet{2024NatSR..1428975O} showed that small uncertainties in the inner boundary of \acrshort{wsa} can propagate into large uncertainties in predicting the solar wind speed.} \citet{McGregor2009} \textcolor{black}{also} showed that improving the source surface boundary, which is used to set the overall orientation of the \acrshort{imf} and dictate solar wind propagation by \acrshort{wsa}, improved the \acrshort{wsa} model skill at predicting both solar wind speed and \acrshort{imf} polarity at Earth. This indicates that even a small error or uncertainty in estimating the orientation of quasi-radial magnetic field lines can significantly affect the prediction of these properties at Earth when utilizing space weather prediction models like \acrshort{wsa}. The ability to quantify and improve this accuracy therefore directly translates into the ability to improve space weather forecasting.

\section{Future Work}

By performing a statistical analysis to determine the similarity of \acrshort{qraft}'s performance between model and white-light data when compared to the expected  \textcolor{black}{\acrshort{mas}} \acrshort{pos B}, we attest that we have defined metric\textcolor{black}{s} \textcolor{black}{that} quantify the accuracy and uncertainty of model outputs compared to white-light \textcolor{black}{\acrshort{pB}} observations. With this framework defined and illustrated, interesting future analyses may be performed. This framework may allow us to explore the latitudinal relationship between the performance of feature tracing and expose regions of the corona that are more poorly traced than others. We can also investigate the performance of the feature tracing analysis based on the length or location of the feature. 

\textcolor{black}{\textcolor{black}{A}nalysis of ground-based observatories such as \acrshort{k-cor} is a planned next step to further validate and analyze this framework. The similar performance of \acrshort{qraft} on both the \acrshort{cor-1} and \acrshort{k-cor} datasets when compared to the expected magnetic orientation would be a significant step in validating this framework between both types of observatories. This section also notes that image artifacts related to background subtraction contributed to error in this analysis. This is of note as
\citet{Thompson2010} compared \acrshort{cor-1} \acrshort{pB} images to Mk4 \acrshort{pB} images to test the effectiveness of background removal, and found the overall behavior to be the same. This is noteworthy as it suggests that the background removal process for each observatory should cause similar error, if any, to the accuracy of the orientation of the quasi-radial features segmented in this study.}

\textcolor{black}{\textcolor{black}{U}sing \acrshort{mas} \acrshort{pos B} as a ground-truth comparison can lead to errors from off-\acrshort{pos} coronal structures introducing signal into the \acrshort{pB} measurements. This error would be worth exploring in future work, and given that \acrshort{mas} models the 3-D structure and morphology of the coronal \acrshort{B}, this could be done by using the \acrshort{los} or emissivity weighted \acrshort{los} integrals of \acrshort{B} components to produce a component in which to compare \textcolor{black}{to}.}

\textcolor{black}{Hydrodynamic boundary conditions and coronal heating models determine the plasma state of the \acrshort{mhd} model \citep[e.g., coronal holes, quiet-sun, active regions;][]{lionello14,downs16,2018NatAs...2..913M}. There is hope that our techniques could be used for heating model parameter optimization in the future, especially if done in tandem with other observational constraints (e.g. EUV and white-light intensities, in-situ measurements, etc.), and we hope to explore this in the future. Additionally, as illustrated in \citet{2018NatAs...2..913M}, which used an ad-hoc method for building shear and magnetic flux-ropes along large-scale polarity inversion lines to construct the energized model under study here, large-scale shear and flux-ropes can influence the morphologies of streamers, particularly at their base. The extent to which the \acrshort{qraft} pipeline is sensitive to such forms of energization is another interesting avenue to explore going forward, but again illustrates the challenge of constraining the large-parameter space involved in coronal modeling, which spans choices made in model parameterizations, physical assumptions, and numerical implementations.}

\textcolor{black}{\textcolor{black}{T}he mean \textcolor{black}{angular difference} was $10.504^\circ$ for \acrshort{ne los} and $12.119^\circ$ for \acrshort{forward pB}. From a theoretical standpoint, one may also make the argument that the \acrshort{pB} scattering should better localize structures to the \acrlong{pos}, as mentioned in \citet{jones2020}, and thus the \acrshort{forward pB} results should perform better than the \acrshort{ne los} results. Further studies may be implemented to test these theories. For example, a cross correlation analysis between the \acrshort{ne los} and \acrshort{forward pB} signal would evaluate the similarity of the 2D arrays for each population, providing an internal check into the validity of this specific result. If further investigation using this framework suggests that the contributions of electrons away from the \acrshort{pos} doesn't produce significant error when comparing traced features to the \acrshort{pos B}, this would be a significant step in validating the central \acrshort{pos} assumption. It should be cautioned that in order to truly validate this assumption, this result would need to be repeated under various conditions and data sources. While this method provides a means in which to do this, it should not be concluded that this lone result constitutes validation of this assumption, and more work is required to definitively come to this conclusion.}

\section{Conclusions}
\label{sec:conclusions}

In this work we have shown several key results:
\begin{enumerate}
    \item We have developed an innovative data analysis framework enabling a quantitative assessment of the performance of image-based coronal magnetic field tracing techniques used in observational solar physics. By studying synthetic coronal images produced by a global \acrshort{mhd} model, we were able to derive a set of rigorous statistical metrics describing the consistency between the image-derived coronal magnetic geometry and the ``ground truth'' coronal magnetic field predicted by the model.

\item The designed methodology has been tested on the \acrshort{qraft} field-line tracing package applied to outputs of the high-resolution \acrshort{mas} model. The results demonstrate that directions of the traced image features are accurate within approximately $\pm10^\circ$ of the \acrshort{pos}-projected magnetic field of the \acrshort{mas} model. This result provides confidence that the large scale magnetic geometry of the \acrshort{pos} projected coronal magnetic field can be measured via remote sensing techniques such as coronal segmentation.

\item In addition to standard statistical metrics such as the average and median \textcolor{black}{\acrlong{angular difference}s} between the segmented features and the coronal magnetic field, we developed a set of entropy-based metrics based on an in-depth analysis of the shape of the misalignment errors characterizing the segmented features. These entropy metrics provide important additional information about how performance profiles compare across different datasets by measuring their similarity.

\item Using the designed ensemble of performance metrics, we evaluated the contribution of several types of errors to the total misalignment error. These errors included image noise, feature-tracing errors, errors introduced by \acrshort{mhd} boundary conditions and first-principle physics included in the coronal model, etc.
\item The introduced methodology has universal application and can be applied to a wide variety of coronal feature tracing methodologies and \acrshort{mhd} models. This research has resulted in a data product available to the space weather community available at \citet{christopher_rura_2025_14921310}.
\end{enumerate}

In conclusion, we developed a set of ready-to-use numerical tools for measuring a geometric discrepancy between a solar coronal model and real coronal structure, ranging from simple statistical indicators to sophisticated distribution comparison metrics. While we attest that this tool will have future application in quantifying the uncertainty in automated feature tracing methods and constraining boundary conditions in space weather models, we show that this tool already has immediate application by quantifying how well an \acrshort{mhd} model of the corona fits the observed structure. We believe this is a valuable framework for space weather modeling, as it provides quantifiable metrics into how well a specific model run fits the global structure of the corona. We \textcolor{black}{also} believe these metrics are valuable to use for coronal segmentations, as they can be used to determine the accuracy of white-light coronal segmentations against a ground-truth solution using an \acrshort{mhd} model. Using this framework, we have shown that a coronal segmentation method identifies the global large-scale orientation of the coronal magnetic field in white-light \acrlong{pB} observations within a reasonable degree of precision.

\facilities{\acrfull{cor-1}}

\software{Astropy \citep{2013A&A...558A..33A}, \acrshort{coronametric} \citep{christopher_rura_2025_14921310}, \acrshort{forward} \citep{2014ascl.soft05007G}, SciPy \citep{2020SciPy-NMeth}, SunPy \citep{sunpy_community2020}}

% \begin{acknowledgments}
We thank the \acrshort{stereo} \acrshort{cor-1} team for providing coronal imaging data, and Predictive Science Inc. for their collaboration and assistance in handling their \acrshort{mas} model solution. C. Rura, S. Jones and V. Uritsky were funded through the Partnership for Heliophysics and Space Environment Research (\acrshort{nasa} grant No. 80NSSC21M0180) and the Windows of the Universe Multi-Messenger Astronomy program (\acrshort{nsf} grant AST-0946422).
C.D. was supported by the \acrshort{nasa} Living With a Star Science (80NSSC22K1021) and Living With a Star Strategic Capabilities (80NSSC22K0893) programs.
N. Alzate acknowledges support from \acrshort{nasa} \acrshort{roses} through HGI grant No. 80NSSC20K1070. C.N.A. and S.I.J were supported in part by the NASA competed Heliophysics Internal Scientist Funding Model (ISFM). C.E.R., S.I.J, and V. M.U. were supported in part by \acrshort{stereo} \acrshort{cor-1} via the PHASER (80NSSC21M0180) and CEPHEUS (NNG11PL10A) awards. 
% \end{acknowledgments}

\printglossary[type=\acronymtype]

\appendix

\section{Gaussian Kernel Density Estimate}
\label{subsec:gaussian_kde}

For a discrete distribution $X_N \equiv \{ X_1, \dots, X_N \}$, a Gaussian kernel density estimate can be defined as:

\begin{equation}
    \hat{f} = \frac{1}{N} \sum_{i=1}^{N} \phi \left( x, X_i; t \right)
\end{equation}

where 

\begin{equation}
    \phi \left( x, X_i; t \right) = \frac{1}{\sqrt{2 \pi t}} e^{-(x - X_i)^2 / 2t}
\end{equation}

note that $N$ is the number of discrete values, and $\phi \left( x, X_i; t \right)$ is a Gaussian kernel that has location $X_i$ and scale $\sqrt{t}$ \citep{2010arXiv1011.2602B}.  This allows us to approximate $N$ different values from an unknown distribution into a continuous probability density function that describes the distribution of values analytically. We note that a bandwidth factor $\hat{h_i}$ can influence the estimate determined by the Gaussian kernel density estimate. We use the bandwidth determined by Scott's rule, which is defined as:

\begin{equation}
    \hat{h_i} = \hat{\sigma_i} n^{-1 / (d+4)}
\end{equation}

where $\hat{\sigma_i}$ is the standard deviation of the data, $n$ is the number of data values, and $d$ is the number of dimensions \citep{2015mdet.book.....S}.

\section{Statistical Metrics of Probability Densities}

\label{subsec:kurtosis_skewness}

Kurtosis is the measurement of whether a dataset is heavy-tailed or light-tailed compared to a normal distribution, while skewness is a measurement of symmetry (or lack thereof) of a dataset \citep{Nist-e-handbook}. These two measurements can be extremely useful in generically describing the overall shape of a given distribution, especially when combined with other statistical measurements such as the mean and standard deviation. 

The following definitions are taken from \cite{Nist-e-handbook}.

For a dataset consisting of $Y_1, Y_2, Y_3, \dots, Y_N$, the kurtosis can be defined as

\begin{equation}
\label{eq:kurtosis}
    kurtosis = \frac{\sum_{i=1}^{N}\left( Y_i - \bar{Y} \right) / N}{s^4} - 3
\end{equation}

Likewise for a dataset consisting of $Y_1, Y_2, Y_3, \dots, Y_N$, the skewness can be defined as 

\begin{equation}
\label{eq:skewness}
    g_1 = \frac{\sum_{i=1}^{N}\left( Y_i - \bar{Y} \right)^3 / N}{s^3}
\end{equation}

where $\bar{Y}$ is the mean of the dataset, $s$ is the standard deviation, and $N$ is the sample size (or number of data points). In expression \ref{eq:kurtosis} subtracting the kurtosis by a factor of three is an alternative definition of kurtosis named the Fisher kurtosis. This is done so that a standard normal distribution has a kurtosis of zero. This is useful for comparing a distribution to the normal distribution, as under these definitions a normal distribution with equivalent mean and standard deviation as the comparison distribution would have zero kurtosis and zero skew.

\section{Jensen-Shannon Divergence}
\label{subsec:JSD_KLD}

The definition of the \acrfull{jsd} is:

\begin{equation}
\label{eq:jsd}
 D_{JS}(P || Q) = \frac{1}{2}D_{KL}(P | | M) + \frac{1}{2}D_{KL}(P | | M)
 \end{equation}
 
 where $P$ and $Q$ are two separate probability distributions, $D_{KL}$ is the \acrfull{kld} which is defined as:
 
 \begin{equation}
 D_{KL}(P || Q) = \sum_x P(x) \log \left(\frac{P(x)}{Q(x)}\right) \label{eq:2}
 \end{equation}
 
 and finally $M$ is a mixture distribution which is defined as:
%   and $M = \frac{1}{2}(P + Q)$ \citep{2020arXiv201012198K}.
 
 \begin{equation}
     M = \frac{1}{2}(P + Q)
 \end{equation}
\citep{2020arXiv201012198K}.

We note from expression \ref{eq:2} that the \acrshort{kld} is defined as the sum of $\log \left(\frac{P(x)}{Q(x)}\right)$. This has the effect of, when comparing a distribution $Q$ to the expected distribution $P$ at $x$, that if $Q > P$, then $\log \left(\frac{P(x)}{Q(x)}\right) >0$, and likewise if $Q < P$, then $\log \left(\frac{P(x)}{Q(x)}\right) <0$. When we sum this expression over $x$, the closer $Q$ represents $P$, the closer the sum approaches $0$. While both \acrshort{kld} and \acrshort{jsd} are metrics for the similarity of two probability distributions, the \acrshort{jsd} is considered the more reliable metric, since the \acrshort{kld} is considered to be asymmetric while the \acrshort{jsd} is a symmetric version of the \acrshort{kld} \citep{2020arXiv201012198K}.

\section{Tukey's Honestly Significant Difference Test}
\label{subsec:Tukey_HSD}

\acrfull{hsd} is a statistical test aimed at measuring the honestly significant difference between two means. \cite{abdi2010tukey} notes that the analysis of variance, or ANOVA, test can indicate that at least one population differs from others in the analysis, but does not describe the pattern of differences between these means. \textcolor{black}{This test aims to determine the \acrfull{H0} that there is no significant difference between two sample populations, with any observed difference being due to sampling or experimental error.}

\acrshort{hsd} aims to calculate the honest statistical difference between these means using what is called the $q$ distribution, which gives the exact sampling distribution of the largest difference between a set of means, $\mu_i - \mu_j$, originating from the same population \citep{abdi2010tukey}. \textcolor{black}{This test provides a \acrfull{ci} for all pair comparisons $i, j ; \ i\neq j$ in which, if $CI_{lower} \leq \mu_i - \mu_j \leq CI_{upper}$ contains $0$ then they are not significantly different, and \acrshort{H0} is accepted. Otherwise all other pairs are significantly different, and \acrshort{H0} is rejected \citep{Nist-e-handbook}. In our analysis, the \acrshort{ci} is chosen at 95\%.} \textcolor{black}{\citet{abdi2010tukey} provides a more rigorous definition of \acrshort{hsd}, and \citet{Nist-e-handbook} provides an example of \acrshort{hsd} that helps to highlight its practical use in statistics.} In our methods, the SciPy routine \textit{scipy.stats.tukey\_hsd} is used to calculate the \acrshort{hsd}.

\bibliography{bibliography}{}
\bibliographystyle{aasjournal}

\end{document}